%% file: main.tex
\documentclass[trackchanges,twocolumn]{aastex631}
\usepackage{amssymb}
\usepackage{float} \usepackage{listings}
\usepackage{xcolor}
\usepackage{amsmath}
\usepackage{hyperref}

\begin{document}
\makeatletter
\let\frontmatter@title@above=\relax
\makeatother

\newcommand\lsim{\mathrel{\rlap{\lower4pt\hbox{\hskip1pt$\sim$}}
\raise1pt\hbox{$<$}}}
\newcommand\gsim{\mathrel{\rlap{\lower4pt\hbox{\hskip1pt$\sim$}}
\raise1pt\hbox{$>$}}}
\newcommand{\CS}[1]{{\color{red} CS: #1}}

\title{\Large Gaia astrometry disfavors a binary origin for long secondary periods}

\shorttitle{Gaia astrometry disfavors a binary origin for long secondary periods}
\shortauthors{Shariat et al.}

\author[0000-0003-1247-9349]{Cheyanne Shariat}
\affiliation{Department of Astronomy, California Institute of Technology, 1200 East California Boulevard, Pasadena, CA 91125, USA}

\author[0000-0002-6871-1752]{Kareem El-Badry}
\affiliation{Department of Astronomy, California Institute of Technology, 1200 East California Boulevard, Pasadena, CA 91125, USA}

\author[0000-0002-1417-8024]{Morgan MacLeod}
\affiliation{Institute for Theory \& Computation, Center for Astrophysics, Harvard \& Smithsonian, Cambridge, MA 02138, USA}

\author[0000-0002-3944-8406]{Emily Leiner}
\affiliation{Department of Physics, Illinois Institute of Technology, 3101 South Dearborn St., Chicago, IL 60616, USA}

\correspondingauthor{Cheyanne Shariat}
\email{cshariat@caltech.edu}

\begin{abstract}
Approximately one-third of luminous pulsating red giant stars exhibit long secondary periods (LSPs): stable photometric variability with periods of several months to years in addition to their much shorter primary pulsation cycles.
Now nearly a century after their discovery, the physical origin of LSPs remains unresolved. A leading explanation invokes binarity, in which the LSP corresponds to the orbital period of a low-mass companion responsible for both the photometric variability and the radial-velocity (RV) modulation. We test this hypothesis using a nearby sample of LSP stars from the {\it Gaia} Focused Product Release, which provides multi-epoch RVs and contemporaneous optical photometry. We find that interpreting the observed RV variability as orbital motion implies companion masses narrowly distributed around $M_2 \approx 0.1~{\rm M_\odot}$ with separations of 1--3 au, placing them squarely in the brown dwarf desert observed around their solar-type progenitors.
Assuming such companions exist, we then forward-model the astrometric signature expected in {\it Gaia} DR3 and predict systematically elevated {\tt RUWE} values for nearby LSPs. In contrast, the observed {\tt RUWE} of nearby LSP stars is systematically lower than these predictions and consistent with most systems exhibiting LSPs being single. This discrepancy disfavors low-mass stellar or substellar companions as the dominant origin of LSPs in evolved stars, motivating a further exploration of alternative stellar mechanisms.
\end{abstract}

\section{Introduction}\label{sec:introduction}

Roughly one in three pulsating red giants show photometric variability on timescales of hundreds to thousands of days, in addition to their dominant pulsation periods of tens to hundreds of days \citep[e.g.,][]{Wood99,Wood04, Soszynski04,Soszynski07,Fraser08,Nicholls09,Pawlak21,Pawlak24}.  
These long secondary periods (LSPs) define a distinct sequence in the period--luminosity ($P$-$L$) diagram \citep[sequence~D;][]{Wood99}.
LSPs have been recognized for nearly a century \citep{OConnell33,PayneGaposchkin54,Houk63}, and although extensive observational and theoretical efforts over this time have characterized many of their empirical properties, their physical origin remains uncertain.

Various mechanisms have been proposed to explain LSPs, including pulsation modes \citep{Wood00,Wood04,Nicholls09,Saio23}, oscillatory convective modes \citep{Saio15,Takayama15,Takayama20,CourtneyBarrer26}, rotational modulation by large convective cells or surface inhomogeneities \citep{Stothers10,Soszynski14}, episodic dust formation and non-spherical mass loss \citep{Wood09,Pawlak21}, and low-mass companions producing orbital modulation and/or dust obscuration \citep{Wood99,Derekas06,Soszynski07,Soszynski21,Goldberg24,MacLeod25,Decin25}.

The past few decades of LSP studies have revealed a number of well-established empirical properties. LSPs remain coherent over many cycles; their light curves are often asymmetric or non-sinusoidal \citep[e.g.,][]{Soszynski14}, their color and effective-temperature variations are modest \citep{Nicholls09,Takayama20}, their radial-velocity amplitudes are only a few km~s$^{-1}$ \citep{Nicholls09}, and they are associated with enhanced mid-infrared excesses relative to non-LSP giants of similar luminosity \citep{Wood09,Pawlak21}. Any successful LSP model must account for this full set of properties \citep[see][for a comprehensive list]{Goldberg24}.

In recent years, binarity has emerged as a leading hypothesis for the origin of LSPs. In this class of models, the LSP corresponds to the orbital period of a low-mass companion -- either a brown dwarf or low-mass star -- embedded in the extended atmosphere or wind of the red giant, while dust associated with the companion or a trailing wake produces the optical and infrared variability \citep[e.g.,][]{Soszynski21}. 
Periodic obscuration by dust associated with the companion, or by a trailing dusty wake, is invoked to explain the optical variability, while thermal emission and eclipses of the dusty material are proposed to explain the infrared behavior \citep{Wood09,Soszynski21,Goldberg24,MacLeod25}. Recent empirical support for this binary picture has come from mid-infrared light curves, which show features resembling secondary minima phased with the LSP, interpreted as eclipses of the dusty structure \citep{Soszynski21}. The binary interpretation has gained further prominence through studies of the nearby supergiant Betelgeuse, whose $\sim2100$~day variability period has been modeled as an LSP caused by a close, low-mass companion \citep[e.g.,][]{Goldberg24,MacLeod25}. Support for this scenario has been derived from radial-velocity, astrometric, X-ray, ultraviolet, and (more tenuously) speckle-imaging constraints on Betelgeuse \citep{Goldberg24,MacLeod25,OGrady25,Howell25,Goldberg25,Dupree26}.

At the same time, the binary hypothesis faces challenges. For example, interpreting the observed LSP radial velocities (RVs) as orbital motion implies companions with masses of roughly $0.05$--$0.15~{\rm M_\odot}$ on separations of order $1$--$3$~au \citep[e.g.,][]{Nicholls09}. This is the regime of the brown dwarf desert, a well-established dearth of companions at these masses and separations around solar-type main-sequence stars, the progenitors of most LSP hosts \citep[e.g.,][]{Marcy00,Grether06}. Reconciling this desert with the high incidence of LSPs, $\sim30\%$ of pulsating red giants \citep{Pawlak21}, is a difficulty for binary-based models \citep[but see also][]{Soszynski21,Decin25}. 

In this work, we present a test of the binary hypothesis for LSPs using a large, homogeneous sample of LSP red giants from the {\it Gaia} DR3 Focused Product Release \citep{Gaia_FPR}. The {\it Gaia} catalog contains $\sim4800$ LSPs with precise RV time series -- nearly two orders of magnitude more than available in previous studies \citep[e.g.,][]{Nicholls09}.
By combining multi-epoch RVs with high-quality photometry, we infer the companion masses implied by a Keplerian interpretation of the LSPs. We then use a forward model of {\it Gaia} astrometry \citep[{\tt gaiamock};][]{ElBadry24} to predict the astrometric signatures expected for such systems and assess whether low-mass companions are present in LSP red giants. 

The remainder of the paper is organized as follows. 
In Section~\ref{sec:methodology} we describe our methodology, including our sample selection and subsequent joint RV+astrometry+photometry analysis. In Section~\ref{sec:results} we present our results and test whether LSPs are consistent with binarity. In Section~\ref{sec:discussion}, we discuss the implications of our results for existing models of LSPs. Our conclusions are summarized in Section~\ref{sec:conclusions}. Supplementary details, figures, and analysis are provided in Appendices \ref{app:gaia_sed_fits}, \ref{app:varying_model}, \ref{app:gaia_dr4}, \ref{app:dust_ruwe_method},  and \ref{app:all_plots}.

\section{Methodology}\label{sec:methodology}
\subsection{LSP Sample Selection}\label{subsec:methodology_sample}
Throughout this paper, we analyze two samples of LSP stars. The first sample is drawn from the {\it Gaia} Focused Product Release (FPR) \citep[][]{Gaia_FPR} and provides RVs, photometry, and astrometry for all systems, enabling a direct test of the binary hypothesis. This serves as the primary sample used throughout the paper.
A secondary sample, drawn from ASAS-SN \citep[][]{Pawlak24}, lacks RVs but increases the number of LSPs available to assess trends that are not heavily reliant on RVs.

\subsubsection{{\it Gaia} Focused Product Release}\label{subsubsec:methodology_sample_FPR}

The main sample used in this work is the {\it Gaia} FPR catalog of long-period variables (LPVs) with RV time series. The catalog provides epoch RVs and contemporaneous multi-band photometry (in $G, G_{\rm BP}, G_{\rm RP}$) for LPV candidates from {\it Gaia} DR3 \citep{Lebzelter23,Gaia_FPR}. The FPR selection is designed to retain only sources with high-quality, well-sampled RV variability with periods consistent with a photometric period. The details of the {\it Gaia} FPR catalog construction are described in \citet{Gaia_FPR}, and we briefly summarize the LSP selection below.

The {\it Gaia} FPR LPV catalog applies a series of stringent quality cuts on the RV data. The cuts include a (i) brightness requirement of $G_{\rm RVS}<12$~mag, (ii) minimum of $12$ RV visibility periods, and (iii) requirement that the uncertainty on the median RV ($\epsilon_{\rm V_R}$) is small compared to the RV amplitude ($\epsilon_{\rm V_R}<0.175\times{\tt rv\_amplitude\_robust}$). Note that ${\tt rv\_amplitude\_robust}$ is the {\it total} peak-to-peak RV amplitude after outlier removal. Additional filtering removes sources with poorly behaved RV time series, including those with more than one rejected RV epoch, significant linear RV trends, or low RV signal-to-noise. After enforcing that the RV-derived period is the same as at least one photometric period (allowing for both $P$ and $2P$), the final FPR catalog contains $9{,}614$ LPVs, of which $6{,}093$ constitute a top-quality subset with fully consistent RV and photometric variability across all bands \citep{Gaia_FPR}. All LSP systems considered in this work are drawn from this top-quality subset. 

LSP stars were selected from the FPR catalog as sources that satisfy \citep[][]{Lebzelter23,Gaia_FPR}:
\begin{equation}\label{eq:gaia_cut}
    A_{\rm G}<0.35~{\rm mag}~~{\rm and}~~ \frac{A_{\rm G}}{{\rm mag}} < 10^{-7} \times \left( \frac{P_{\rm G}}{{\rm days}}\right)^{2.5},
\end{equation}
where $A_{\rm G}$ is the photometric $G$-band semi-amplitude and $P_{\rm G}$ is the corresponding photometric period. Filtering for low-amplitude, long-period variables effectively excludes the majority of fundamental-mode pulsators \citep[][]{Gaia_FPR}.
For the analysis throughout this paper, we further restrict the sample to nearby stars ($d \leq 1.5$~kpc; $N=224$). At distances larger than $\sim1$~kpc, the astrometric wobble due to low-mass companions at few-au separations will typically be too small to be detected by {\it Gaia} and to inflate the {\tt RUWE} value \citep[][]{Belokurov20, Sullivan25, ElBadry25}.

\subsubsection{ASAS-SN variable stars}
\label{subsec:methodology_asas_sample}

We also consider an independent sample of LSP stars identified from ASAS-SN photometry \citep{Pawlak24}, which only minimally overlaps with the {\it Gaia} FPR sample (8 common stars). These systems do not have epoch RVs; thus, companion masses cannot be derived directly. Instead, we assign simulated companion masses by sampling from the {\it Gaia} LSP distribution, which is narrowly peaked at $M_2\approx0.1~{\rm M_\odot}$ (see Section \ref{subsec:results_rvamps}).

\citet{Pawlak24} selected LSP stars from the ASAS-SN Catalog of Variable Stars \citep{Jayasinghe18, Jayasinghe19} and cross-matched them with {\it Gaia} DR3. They identified LSPs using an empirical period--amplitude criterion designed to isolate stars on ``sequence~D'' \citep[e.g.,][]{Wood99} while avoiding contamination from long-period pulsators and ellipsoidal binaries. For each star, the three strongest periods were identified, and a period was classified as an LSP if it satisfied
\begin{equation}\label{eq:asas_selection}
    \frac{A_V}{\rm mag} < 1.6~\log \left( \frac{P_V}{\rm days} \right) - 3.7,
\end{equation}
where $P_V$ is the photometric period in days and $A_V$ is the peak-to-peak $V$-band amplitude. This criterion is calibrated using OGLE Large Magellanic Cloud LPVs and was demonstrated to select stars occupying the classical LSP sequence in the $P$-$L$ diagram \citep{Pawlak24}. An additional magnitude-dependent amplitude cut, $A > 0.036V - 0.458$, was imposed on faint stars to mitigate spurious detections, where ASAS-SN photometric uncertainties increase ($\approx 0.1$~mag at $V=17$~mag).

From the \citet{Pawlak24} LSP catalog ($N=55{,}558$), we choose stars with positive {\it Gaia} parallaxes and $d<1.5$~kpc ($N=1{,}601$). The distance cut is a conservative threshold for resolving astrometric motion induced by low-mass companions in {\it Gaia} DR3 \citep[][]{ElBadry24}; the photocenter of a hypothetical LSP binary at $1.5$~kpc is $a_0\lsim0.2$~mas. To exclude residual main-sequence contaminants, we require an absolute {\it Gaia} $G$-band magnitude $M_G < 1$ ($N=1{,}370$). Finally, to avoid very bright sources that are likely affected by saturation or unreliable ASAS-SN photometry, we impose a cut of $G > 5$, leaving a final sample of $N=1{,}362$ red giant LSPs.

\subsubsection{Summary of the Sample}

\begin{figure}
\centering
\includegraphics[width=1.0\columnwidth]{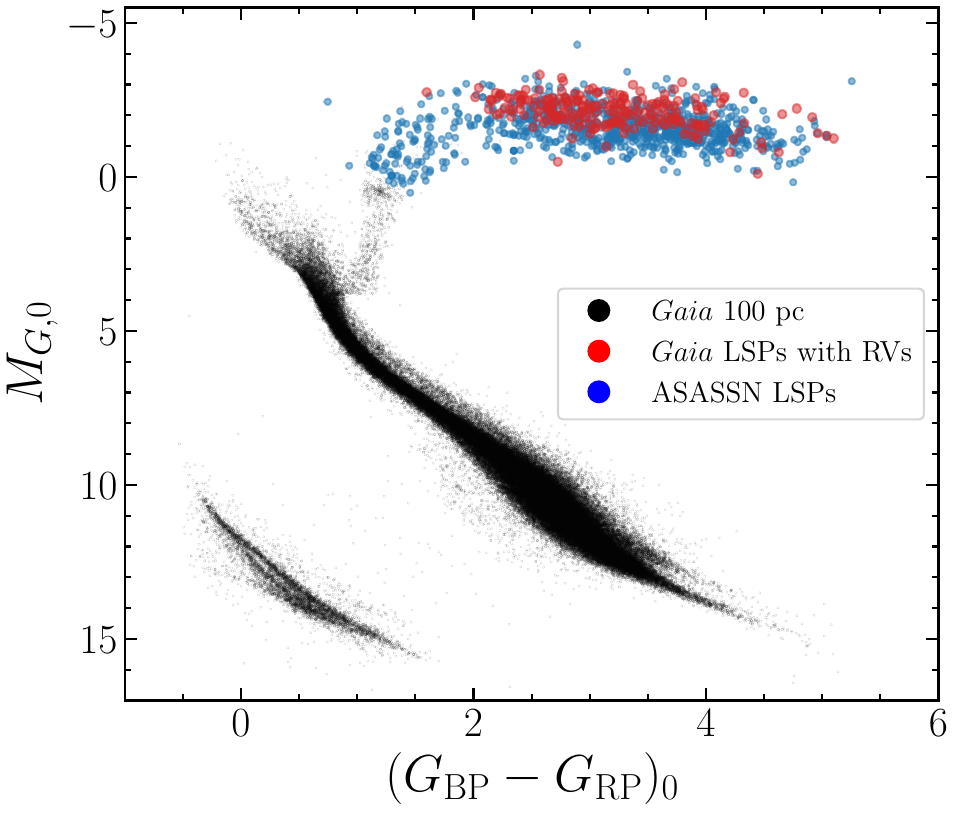}
\caption{Color-magnitude diagram of the LSP samples used in this work. 
We show $1.5$~kpc LSPs from the {\it Gaia} Focused Product Release with RV time-series (red), 
which constitutes the primary sample used in this work, as well as the ASAS--SN LSP sample (blue). 
The cleaned {\it Gaia} 100 pc sample is shown for comparison (black). All {\it Gaia} photometry is dust-corrected.} \label{fig:hr_diagram_lsps}
\end{figure}

Figure \ref{fig:hr_diagram_lsps} shows the location of the {\it Gaia} and ASAS--SN LSP samples on the dust-corrected {\it Gaia} CMD. 
Extinction corrections are derived using the three-dimensional dust map of \citet{Edenhofer24}, which reconstructs the interstellar dust distribution from {\it Gaia} XP spectra.
Most systems are evolved RGB and AGB stars, often near or above the tip of the RGB, consistent with previous findings \citep[e.g.,][]{Pawlak21}. In the $d<1.5$~kpc sample, the two catalogs overlap along the evolved giant branch, but ASAS--SN extends to bluer colors and slightly fainter magnitudes on the earlier RGB, whereas the {\it Gaia} FPR sample does not. The plotted samples contain $895$ ASAS--SN LSPs and $224$ {\it Gaia} LSPs, and the clearest mismatch is the blue extension at $(G_{\rm BP}-G_{\rm RP})_0 \lesssim 2.2$, where ASAS--SN contains $131$ systems and the {\it Gaia} FPR contains only $9$.

These blue ASAS--SN sources trace a lower-luminosity subset of the nearby LSP population. They are fainter in apparent magnitude than the stars in the shared evolved giant-branch region, with median $G=10.49$, compared to $G=9.19$ for the remaining ASAS--SN systems and $G=7.91$ for the {\it Gaia} sample in the same distance range. They also show somewhat shorter LSPs, with median $P_{\rm LSP}\simeq424$~d compared to $566$~d for the redder ASAS--SN systems.

Their absence from the {\it Gaia} sample is best explained by the RV-based selection of the {\it Gaia} FPR. The blue ASAS--SN systems are concentrated toward the low-latitude disk, with $61\%$ at $|b|<10^\circ$ and $39\%$ at $|b|<5^\circ$, compared to $28\%$ and $16\%$ for the {\it Gaia} sample. Within the {\it Gaia} FPR sample, the main CMD trend is a decline in RV signal-to-noise toward fainter systems, while the measured RV amplitudes vary only weakly across the occupied locus. The blue ASAS--SN extension therefore reflects a nearby LSP population that is recovered efficiently in photometry but is underrepresented in the {\it Gaia} RV-selected sample, especially at lower luminosities and low Galactic latitude.

\begin{figure}
\centering
\includegraphics[width=1.0\columnwidth]{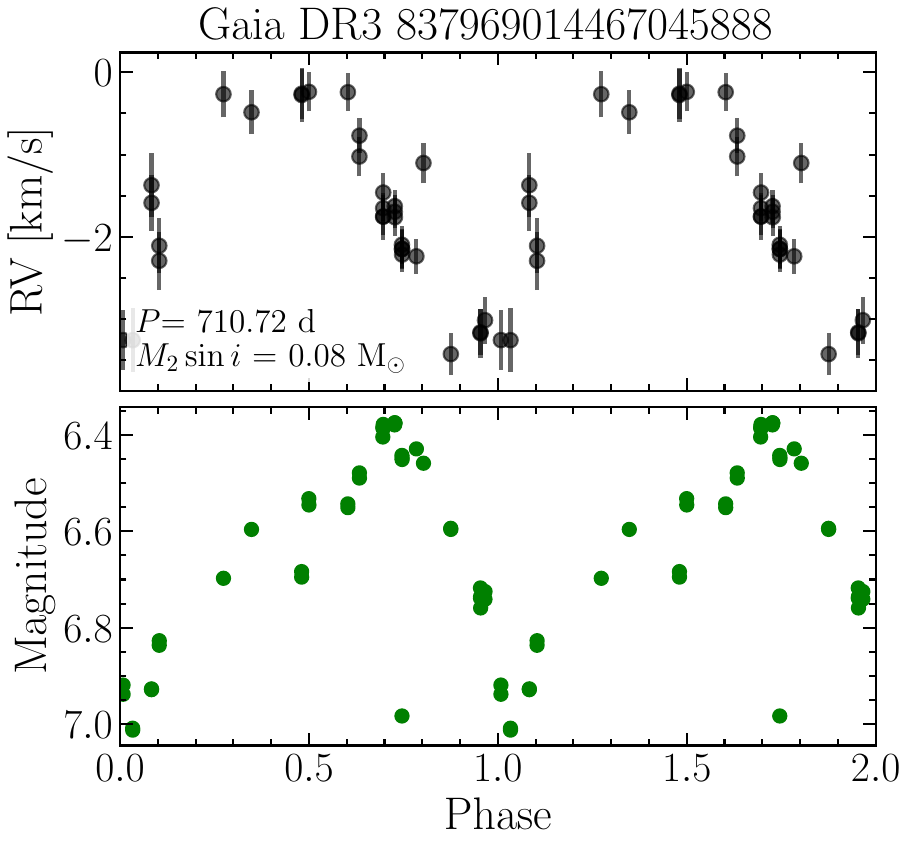}
\caption{Example of RV and photometric time-series observations of LSPs in the {\it Gaia} FPR sample. We show measured RVs (top panel) and $G$-band photometry (bottom panel) for an example LSP: Gaia DR3 837969014467045888. We include both the time series (left) and phase-folded data (right), folded on the long secondary period.
The star resides at a distance of $683$~pc with a long secondary period of $711$~days. If the RV variations were due to orbital motion about a $1~{\rm M_\odot}$ primary, the secondary's minimum mass would be $M_2 \sin i=0.08~{\rm M_\odot}$. For the phase-folded time-series of every {\it Gaia} LSP within $750$~pc, see Appendix \ref{app:all_plots}.
}
\label{fig:example_rv_phot}
\end{figure}

Figure \ref{fig:example_rv_phot} shows an example of the RV and photometric time series observations provided by the {\it Gaia} FPR. The chosen system, Gaia DR3 837969014467045888, exhibits a long secondary period of $711$ days at a distance of $683$~pc. The phase-folded figure reveals a phase offset of $+0.2$ between the RV and photometric time series, consistent with previous results on LSPs \citep[e.g.,][]{Nicholls09,Goldberg24}.
The RV semi-amplitude of $1.48~{\rm km~s^{-1}}$ corresponds to an implied minimum companion mass of $M_2 \sin i=0.08~{\rm M_\odot}$, assuming the RVs are due to orbital motion about a $M_1=1~{\rm M_\odot}$ primary star. These features -- phase offset, period, and RV amplitude -- are broadly similar to the rest of the {\it Gaia} LSP sample (see Appendix \ref{app:all_plots}).

\subsection{Mock {\it Gaia} astrometry}\label{subsec:methodology_gaiamock}

To assess whether the companion masses implied by a binary interpretation of LSPs are compatible with {\it Gaia} astrometric constraints, we generate mock {\it Gaia} observations for each source within $1.5$~kpc (see Section \ref{subsec:methodology_sample}) assuming they are binaries with companion masses derived from RVs. Mock {\it Gaia} observations are simulated for each hypothetical binary using the \texttt{gaiamock} package\footnote{\url{https://github.com/kareemelbadry/gaiamock}} \citep{ElBadry24}. The code simulates epoch astrometric measurements following the {\it Gaia} scanning law and models the effects of orbital motion on the photocenter using the formalism of \citet{Lindegren22}. Measurement uncertainties are assigned using an empirical noise model calibrated on the residuals of well-behaved sources in {\it Gaia} DR3 \citep{Holl23}.

The simulated epoch astrometry is processed through the same cascade of astrometric models employed in the {\it Gaia} DR3 pipeline, including single-star and non-single-star solutions, following the procedures described by \citet{Halbwachs23}. This yields a synthetic astrometric solution for each mock system, including goodness-of-fit statistics that can be directly compared to those reported in the {\it Gaia} DR3 catalog \citep[][]{ElBadry24, ElBadry25}.

A key diagnostic in {\it Gaia} is the re-normalized unit weight error ({\tt RUWE}), which quantifies the quality of the single-star astrometric fit.
By construction, well-modeled single stars have ${\tt RUWE}\approx1$ \citep{Lindegren18}. For unresolved binaries, orbital motion of the photocenter introduces systematic residuals that inflate the {\tt RUWE}, with the magnitude of the effect depending on the orbital parameters, sky position, and apparent magnitude of the system. As a result, {\tt RUWE} provides a sensitive and quantitative diagnostic of unresolved binarity in {\it Gaia} data, and its expected value can be robustly predicted for a given binary configuration using {\tt gaiamock} \citep[e.g.,][]{ElBadry24,ElBadry25}.

In this work, we use a modified version of {\tt gaiamock} designed to improve the reliability of {\tt RUWE} predictions, particularly for systems with small photocenter orbits and for effectively single stars \citep{Iorio26}. The default version of the code accurately reproduces the {\tt RUWE} distribution for binaries with clearly detectable astrometric motion \citep[e.g.,][]{ElBadry25}, but systematically underestimated the width of the {\tt RUWE} distribution at low signal-to-noise. The modified version of {\tt gaiamock}\footnote{also publicly available in the {\tt gaiamock} GitHub repository} used here eliminates CCD-level binning and instead treats individual CCD measurements explicitly. In addition, it applies an empirical, sky-position-dependent rescaling of the epoch-level astrometric uncertainties to reproduce the observed {\tt RUWE} distribution of single stars in {\it Gaia} DR3.

\section{Results}\label{sec:results}

\subsection{RV amplitudes and companion masses}\label{subsec:results_rvamps}

\begin{figure*}
\centering
\includegraphics[width=1.0\textwidth]{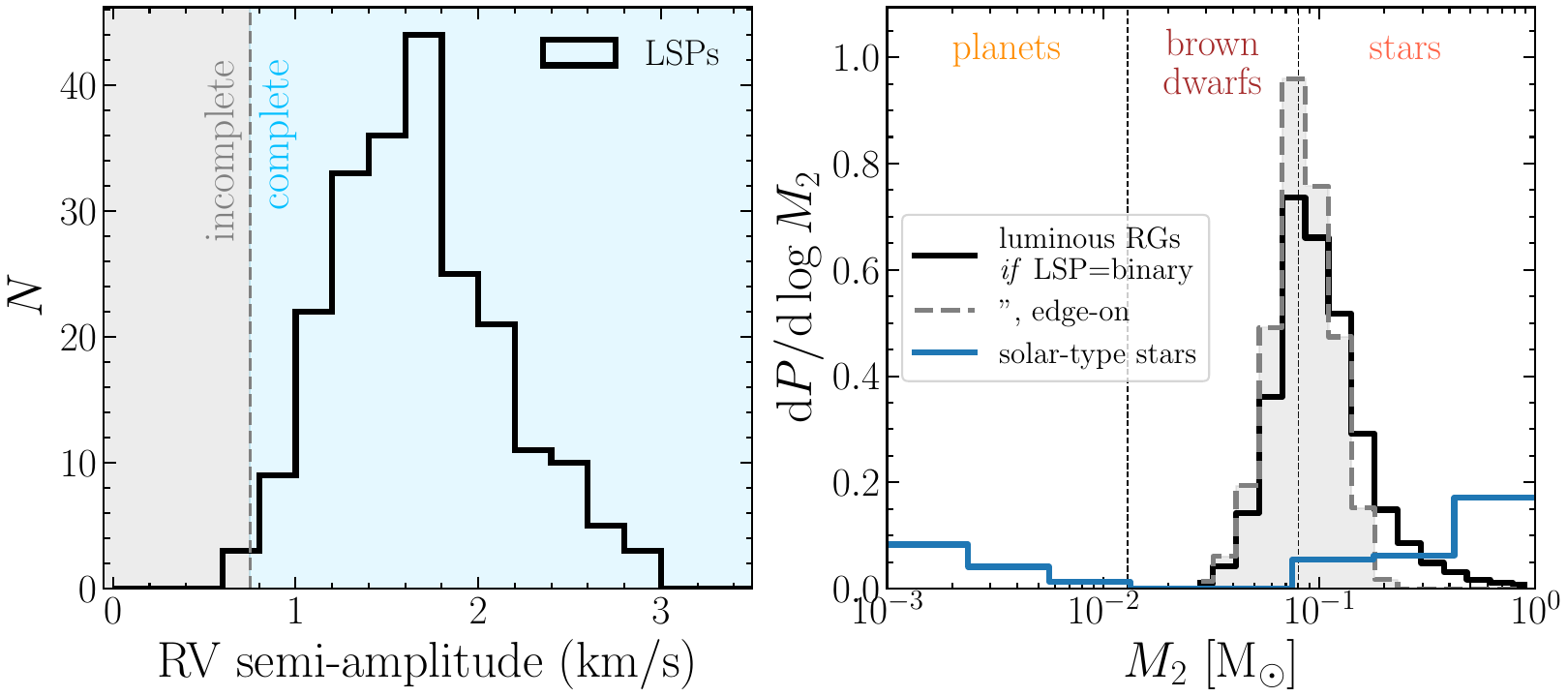}
\caption{RV semi-amplitude (left) and companion mass (right) probability distribution of LSPs in the {\it Gaia} $1.5$~kpc sample ($N=224$). 
On the left, we show the $K=0.75~{\rm km~s^{-1}}$ line, above which our sample is roughly complete. On the right, the companion mass, $M_2$, is derived from $K$ assuming a solar-mass primary and circular orbit. We show the result for both an isotropic inclination distribution (solid black) and one restricted to edge-on inclinations $45^\circ \lesssim i \lesssim 135$ (gray-dashed). We compare LSPs to the $M_2$ distribution of solar-type stars with $M_1 \geq 1~{\rm M_\odot}$ \citep[blue solid;][]{Grether06}. Solar-type stars show a dearth of $0.02-0.1~{\rm M_\odot}$ companions (the ``brown dwarf desert''), whereas luminous red giants ($\sim30\%$ of which are LSPs) suggest ubiquitous $0.02-0.1~{\rm M_\odot}$ companions {\it if} binarity were the cause of LSPs.}
\label{fig:rv_amps_m2s}
\end{figure*}

For each LSP star, we use the observed RV variability to infer the companion mass implied by assuming it is due to Keplerian motion. We adopt a fiducial primary mass of $M_1 = 1~{\rm M_\odot}$ and compute the RV mass function,
\begin{equation}\label{eq:rv_mass_func}
f(M) \equiv \frac{(M_2\sin i)^3}{(M_1+M_2)^2} = \frac{P K^3}{2\pi G}~(1-e^2)^{3/2},
\end{equation}
where $P$ is the LSP period, $K$ is the RV semi-amplitude, $i$ is the orbital inclination, and $e$ is the eccentricity. 
We assume circular orbits for these estimates and consider both isotropic and edge-on inclination distributions. The isotropic case provides the default expectation for a randomly oriented population, while the edge-on case is motivated by models where the observed variability is due to eclipses\citep[e.g.,][]{Soszynski21, Decin25}. Appendix \ref{app:varying_model} shows the results for varying assumptions.

Figure \ref{fig:rv_amps_m2s} shows the distribution of RV semi-amplitudes (left) and the corresponding companion-mass estimates (right) for the $1.5$~kpc {\it Gaia} LSP sample ($N=224$). The observed RV semi-amplitudes are clustered sharply with median $K = 1.63~{\rm km~s^{-1}}$ and standard deviation $~\sigma_K = 0.5~{\rm km~s^{-1}}$. Assuming that the RV variability is due to orbital motion and adopting a fiducial primary mass of $M_1 = 1~{\rm M_\odot}$, these amplitudes translate into a sharply peaked companion-mass ($M_2$) distribution.
The inferred secondary masses peak near the hydrogen-burning limit, with a median $M_2 \approx 0.084 ~{\rm M_\odot}$ and standard deviation $\sigma_{M_2}=0.086~{\rm M_\odot}$. The edge-on distribution is even narrower, with a similar peak but only $\sigma_{M_2}=0.03~{\rm M_\odot}$.

A striking feature of the RV amplitude distribution is the paucity of low-amplitude systems. In the $1.5$~kpc {\it Gaia} sample, only $5\%$ of stars have $K<1~{\rm km~s^{-1}}$. Part of this deficit is due to selection effects of the {\it Gaia} FPR sample. In particular, the requirement $\epsilon_{\rm V_R}<0.175\times{\tt rv\_amplitude\_robust}$ implies that the RV semi-amplitude, ${\tt rv\_amplitude\_robust}=2K$, must exceed approximately three times the uncertainty on the median RV. For the sample analyzed here, the distribution of median RV uncertainties peaks near $0.25~{\rm km~s^{-1}}$ (standard deviation $0.13~{\rm km~s^{-1}}$), such that systems typically require $K\gtrsim0.7$--$0.8~{\rm km~s^{-1}}$ to be included.

However, selection effects alone cannot explain the observed distribution. Among the $224$ LSP stars, $57\%$ have $\epsilon_{\rm V_R}<0.25~{\rm km~s^{-1}}$, implying that they are sensitive to RV semi-amplitudes down to $K\approx0.75~{\rm km~s^{-1}}$. Despite being sensitive to low $K$ values, only $12$ ($5\%$) of these systems exhibit $K<1~{\rm km~s^{-1}}$. Namely, the RV semi-amplitudes {\it intrinsically} cluster tightly around $K\sim1$--$2~{\rm km~s^{-1}}$, with relatively few systems at $K<1~{\rm km~s^{-1}}$ despite sensitivity there. This is shown by the shaded regions in Figure \ref{fig:rv_amps_m2s}, where the sample is roughly complete in the blue region.

This conclusion is consistent with the independent results of \citet{Nicholls09}, who obtained multi-epoch RV measurements for 58 LSP red giants using the FLAMES/GIRAFFE spectrograph on the VLT \citep{Pasquini02} at a resolving power of $R\approx23{,}900$ over $693.7$--$725.0$~nm. They likewise found a narrow peak in RV semi-amplitudes at $K\approx1.75~{\rm km~s^{-1}}$ and a pronounced deficit of systems below $K\sim1~{\rm km~s^{-1}}$. Through injection--recovery simulations, they demonstrated that their data were sensitive to RV semi-amplitudes below $1~{\rm km~s^{-1}}$, yet detected none, leading them to conclude that the intrinsic RV amplitude distribution of LSP stars is sharply peaked near $K\sim1$--$2~{\rm km~s^{-1}}$. Our results independently confirm this conclusion using the larger sample of {\it Gaia} RVs.

The right panel of Figure \ref{fig:rv_amps_m2s} shows the companion mass distribution of LSPs as inferred from their RV amplitudes, assuming isotropic line-of-sight inclinations and solar mass primaries. We calculate the probability of a luminous red giant (RG) hosting a companion in the $1-3$~au range per logarithmic range of secondary mass ($N_i$) for a bin of width $\Delta \log M_2$ using
\begin{equation}\label{eq:dn_dlogm}
    \frac{{\rm d}P}{{\rm d}\log M_2} = \frac{N_i}{N_{\rm tot}\Delta \log M_2}.
\end{equation}
Because the right-panel LSP curve is normalized per luminous RG rather than per LSP host, its integral is $f_{\rm LSP} \approx 0.3$ \citep[e.g.,][]{Wood99,Wood04,Pawlak21}.
We also calculate this quantity for the $25$~pc sample of solar-type stars with $M_1 \geq 1~{\rm M_\odot}$ \citep[][]{Grether06}, which are the progenitors of luminous RGs. The \citet{Grether06} sample contains $464$ Hipparcos stars, $384$ of which have RVs. They characterize the completeness-corrected distribution of ${\rm d}N/{\rm d}\log M_2 = N_i/\Delta \log M_2$ for companions more massive than Jupiter --  including giant planets, brown dwarfs, and stars -- with orbital periods shorter than $5$~years. Overall, they find that $\sim16\%$ of solar-type primaries have companions in this mass and period range: $\sim5\%$ are giant planets, $\sim11\%$ are stars, and $<1\%$ are brown dwarfs.

The inferred companion-mass distribution for LSP stars is strongly concentrated near the stellar--substellar boundary (Figure~\ref{fig:rv_amps_m2s}). Correcting for isotropic inclinations gives a posterior median of $M_2\approx0.09~{\rm M_\odot}$ with $\sigma=0.045~{\rm M_\odot}$. Roughly one-third ($35\%$) of the systems are consistent with brown-dwarf-mass companions ($M_2 < 0.08~{\rm M_\odot}$) and $82\%$ are consistent with $M_2 < 0.15~{\rm M_\odot}$.

Interpreted as orbital motion, the RV variability of LSPs thereby implies companions with masses of $0.03$-$0.15~{\rm M_\odot}$ with orbits at $\sim1$-$3$~au around evolved solar-type stars (Section~\ref{subsec:results_rvamps}). 
For solar-type stars on the main-sequence, such companions are notoriously rare, defining the ``brown dwarf desert,'' a pronounced deficit of companions with masses $\sim0.02$-$0.2~{\rm M_\odot}$ at separations of a few au around solar-type main-sequence stars \citep[e.g.,][]{Marcy00,Grether06}. This can be appreciated from the $M_2$ distribution of solar-type main-sequence stars shown in Figure \ref{fig:rv_amps_m2s} (blue curve). 
Fewer than $1\%$ of solar-type main-sequence stars have companions in this mass and period range $0.03$-$0.15~{\rm M_\odot}$ whereas $\sim30\%$ of RGs of similar mass show LSPs, about half of which lie in the mass range \citep[][]{Soszynski21,Pawlak24}. Reconciling the high incidence of LSPs ($\sim30\%$) among luminous red giant and AGB stars with the low incidence of brown-dwarf companions orbiting their main-sequence progenitors is a fundamental challenge for binary-based models. 
Furthermore, if such companions exist and survive post-main-sequence evolution, a comparable fraction of white dwarfs should host $\sim0.1~{\rm M_\odot}$ companions at separations of a few au. However, IR excess surveys find brown dwarf companions around $\lesssim1\%$ of white dwarfs \citep[e.g.,][]{Farihi05,Girven11,Steele11}.

One proposed resolution is that many LSP companions did not form at their present masses but were initially planetary-mass companions that grew through wind accretion or mass transfer as the host ascends the giant branch, evolving into brown dwarfs or very low-mass stars \citep[e.g.,][]{Retter05,Soszynski21}. This scenario predicts that LSPs presently reside in binaries with low-mass companions, which would induce astrometric motion to LSPs as observed by {\it Gaia}. We test this prediction in Section \ref{subsec:results_ruwe}.

\subsection{Gaia {\tt RUWE}}\label{subsec:results_ruwe}
To test the binary interpretation of LSPs, we compare the astrometric signal predicted from the RV-inferred orbits to the observed {\it Gaia} astrometry. Assuming that the RV variability is from a companion's orbit, we forward-model the corresponding {\it Gaia} DR3 astrometric observations and compare them to the observed {\tt RUWE} values.

We forward-model the {\it Gaia} astrometric response expected for these hypothetical binaries using {\tt gaiamock} \citep{ElBadry24}. For each star, we construct a binary model with period fixed to the observed LSP, minimum companion mass ($M_2 \sin i$) fixed to the RV-inferred value, and primary mass fixed to $1~{\rm M_\odot}$. We assume a dark companion that contributes negligible flux compared to the evolved star, such that the photocenter follows the primary to an excellent approximation\footnote{In the optical {\it Gaia} $G$ band, a typical LSP red giant is $\sim10^4$ times brighter than even the highest-mass M-dwarf companion, so the secondary light is neglected when computing the photocenter orbit}. We sample orbital inclinations isotropically (uniform in $\cos i$) and draw the longitude of the ascending node $\Omega$ and argument of periastron $\omega$ from uniform distributions $\mathcal{U}(0,2\pi)$. For each star, we generate $50$ Monte Carlo realizations, each time (a) drawing angles, (b) running {\tt gaiamock} to produce synthetic epoch astrometry at the star's sky position and brightness, and (c) recording the resulting {\tt RUWE} from the best-fit single-star solution. Our fiducial models assume $e=0$ and isotropic inclinations. 
However, it has been proposed that LSP companions preferentially reside in edge-on orbits that are eccentric \citep{Nicholls09}. Thus, we also re-simulate the astrometry assuming different assumptions on inclination and eccentricity, finding that our conclusions remain unchanged (Appendix \ref{app:varying_model}). For each star, we compare the observed {\tt RUWE} from {\it Gaia} DR3\footnote{a DR4 analogue of this test is shown in Appendix~\ref{app:gaia_dr4}} to the distribution of {\tt RUWE} values predicted across Monte Carlo realizations, thereby testing whether the RV-inferred companions are compatible with {\it Gaia} data.

\begin{figure}
\centering
\includegraphics[width=1.0\columnwidth]{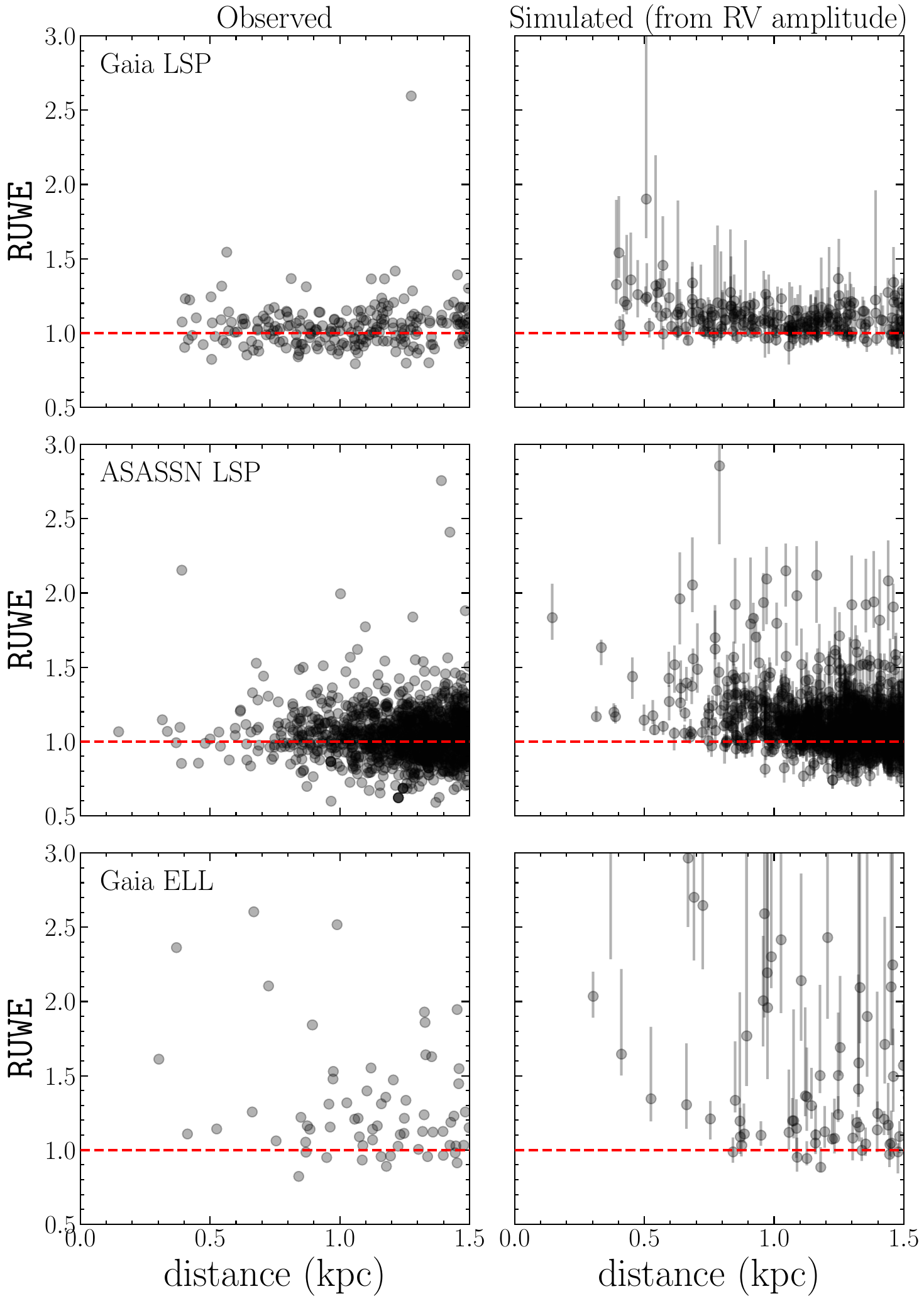}
\caption{Observed vs predicted {\tt RUWE} for LSP stars assuming the binary hypothesis. 
The left panel shows the observed {\it Gaia} {\tt RUWE} while the right panel shows the predicted {\tt RUWE} assuming the RV variability is due to a binary companion. We display results for three different samples, restricted to $1.5$~kpc: {\it Gaia} LSPs (top), ASAS--SN LSPs (middle), and {\it Gaia} ellipsoidal variables (ELL; bottom). Note that the ASAS--SN sample does not have RVs. The red dashed line shows ${\tt RUWE}=1$. 
The nearest LSPs are predicted to show inflated {\tt RUWE} values if the LSP was due to binarity; however, they do not. In contrast, the nearest ellipsoidal variables are predicted to show inflated {\tt RUWE} values if the RV variability is due to binarity, which they do.
}
\label{fig:RUWE}
\end{figure}

Figure \ref{fig:RUWE} compares the {\tt RUWE} values predicted under the binary hypothesis to those measured in {\it Gaia} DR3. The left column shows the observed {\tt RUWE} as a function of distance from {\it Gaia} DR3, while the right column shows predictions assuming LSPs reside in binaries. The error bars reflect $1\sigma$ uncertainties ($16$th-$84$th percentile).
We show the results for three samples: (1) {\it Gaia} LSPs, (2) ASAS--SN LSPs, and (3) {\it Gaia} ellipsoidal variables (ELL), also selected from the {\it Gaia} LPV sample of \citep[][]{Gaia_FPR}. The {\it Gaia} ELL sample serves as a control to demonstrate that {\tt RUWE} closely tracks binarity, particularly in the nearest systems ($d\lsim1.5$~kpc).

If the LSPs were due to orbiting companions with masses implied by their RVs (often $\approx0.1~{\rm M_\odot}$; see Figure \ref{fig:rv_amps_m2s}), the binary motion would be detected by {\it Gaia} and lead to inflated {\tt RUWE} values for the closest systems (top two panels). However, this is not observed. Instead, the majority of LSPs within $750$~pc ($80\%$, $58/73$) show predicted {\tt RUWE} larger than observed, indicating that the astrometric motion expected from the companion is not present in the {\it Gaia} data. A small number of systems do show elevated {\tt RUWE} values consistent with unresolved binaries within 750 pc: Gaia DR3 1046797436863973504 (${\tt RUWE}=1.54$) and Gaia DR3 4933972912553085440 (${\tt RUWE}=1.3$). Nevertheless, the LSP population in aggregate does not show the strong increase in {\tt RUWE} toward smaller distances predicted if the majority of LSPs were binaries (top-right panel), indicating that such cases represent a minority of the population.
The discrepancy between observed and predicted {\tt RUWE} persists out to $d\lesssim750$~pc, where the astrometric reflex from $0.1~{\rm M_\odot}$ companions is detectable in {\it Gaia} DR3. 
In contrast, the {\it Gaia} ellipsoidal variable candidates, which typically have $K\approx10~{\rm km~s^{-1}}$ ($M_2\approx0.65~{\rm M_\odot}$), do show the expected elevated {\tt RUWE}: within 750 pc, $78\%$ ($7/9$) have observed ${\tt RUWE}>1.2$, while the forward model predicts ${\tt RUWE}>1.2$ for all nine systems. This is consistent with {\it Gaia} observations and with them being genuine binaries.

As a complementary assessment, we also perform a two-sample KS test comparing the observed {\tt RUWE} distribution of LSPs within 1 kpc to the distributions predicted by the forward model. All of the binary models we considered, including variants with eccentric orbits and non-isotropic inclinations (Appendix \ref{app:varying_model}), yield $p<5\times10^{-2}$. For comparison, the same test applied to the nearby ELL control sample within 750 pc gives $p=0.73$, suggesting that the observations {\it are not} discrepant with the model.
Note that we do not rely on the KS test in our main argument. The central result is the population-level trend visible in Figure \ref{fig:RUWE}, namely that the binary hypothesis predicts systematically increasing {\tt RUWE} for the nearest LSPs, while the observed {\tt RUWE} remain near unity and show no such rise.

In summary, if LSPs were caused by binary companions with masses inferred from their RV amplitudes ($0.03-0.2~{\rm M_\odot}$), the resulting orbital motion should produce detectable astrometric signatures and inflate the {\it Gaia} {\tt RUWE} of nearby systems. Instead, we find that the observed {\tt RUWE} of LSPs under the binary hypothesis is significantly smaller than the predicted value, indicating that the expected astrometric reflex motion is not present. By contrast, known ellipsoidal-variable binaries show elevated {\tt RUWE} values in {\it Gaia}, which are consistent with what is predicted from their RV amplitudes with the forward model (Figure \ref{fig:RUWE}). We therefore conclude that (sub)stellar companions with masses implied by the RV variability are unlikely to be the dominant cause of long secondary periods in red giants.

\subsection{Comparing LSP pulsators to non-LSP pulsators}\label{subsec:comp_LSPs_nonLSPs}

To assess whether LSP stars are astrometrically unusual compared to other luminous pulsating giants, we construct a control sample from the same {\it Gaia} FPR LPV-RV parent catalog used in the main analysis. We define non-LSP pulsators as sources that satisfy the pulsation criteria \citep[see figure 11 in][]{Gaia_FPR} and fail the LSP selection in Equation~\ref{eq:gaia_cut}. We also exclude likely ellipsoidal systems using the FPR amplitude--RV cut, and restrict the control sample to the same absolute-magnitude and distance range ($d<1.5$~kpc) as the LSP sample \citep{Gaia_FPR}. 

\begin{figure*}
\centering
\includegraphics[width=0.99\textwidth]{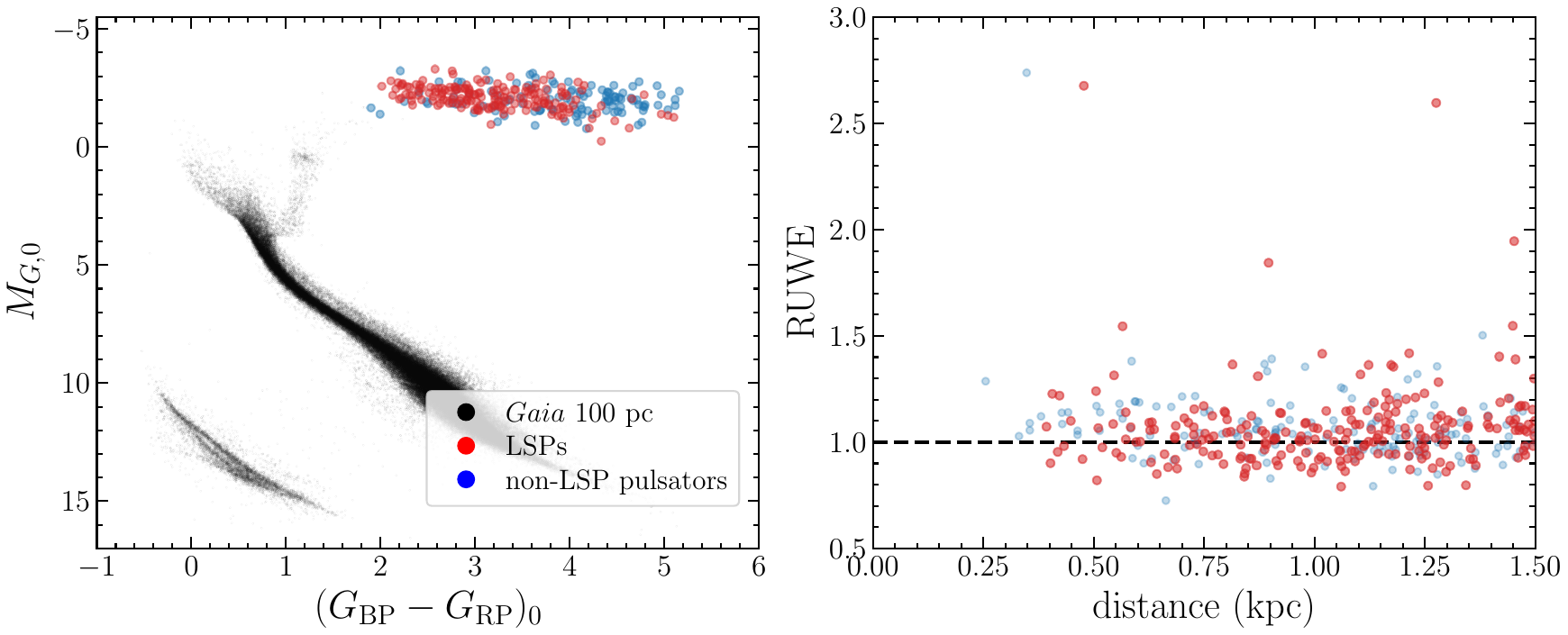}
\caption{ Comparison of LSP and non-LSP pulsators in the {\it Gaia} FPR sample. {\it Left:} dust-corrected {\it Gaia} CMD for nearby LSPs (red) and non-LSP pulsators (blue), shown together with the cleaned {\it Gaia} 100 pc sample (black) for reference. {\it Right:} observed {\tt RUWE} versus distance for the same LSP and non-LSP pulsator samples. Both LSP and non-LSP pulsators show similar {\tt RUWE} values concentrated near ${\tt RUWE}\approx1$, disfavoring a model where LSP pulsators preferentially reside in binaries.}\label{fig:lsp_nonlsp_cmd_ruwe}
\end{figure*}

Figure~\ref{fig:lsp_nonlsp_cmd_ruwe} compares the location of the two samples on the dust-corrected {\it Gaia} CMD (left) and their observed {\tt RUWE} as a function of distance (right). The two populations occupy broadly similar regions of the evolved giant branch, although the non-LSP pulsators extend somewhat redward at the bright end and show a slightly broader luminosity distribution.
The observed {\tt RUWE} distributions are also similar. In particular, the LSP sample does not show a clear excess of large {\tt RUWE} values relative to non-LSP pulsators of comparable luminosity and distance, and both populations remain clustered near ${\tt RUWE}\approx1$ over most of the distance range shown. In fact, the median observed ${\tt RUWE}$ for non-LSP pulsators ($1.12$) is larger than that for LSP pulsators ($1.03$) at close distances ($d<750$~pc). 
This comparison supports the main conclusion of Section~\ref{subsec:results_ruwe}: LSP stars do not behave astrometrically like a population dominated by unresolved binaries with low-mass companions, but instead resemble other pulsating RGB/AGB stars in their overall single-star astrometric goodness of fit.

\section{Discussion}\label{sec:discussion}

\subsection{The effect of dust obscuration and pulsations on astrometry}\label{subsec:dust_discussion}

One possible consideration is that LSP-related dust obscuration or intrinsic stellar variability could shift the optical photocenter, thereby affecting the {\it Gaia} astrometry. To test this, we used {\tt gaiamock} to inject photocenter motions into synthetic single-star DR3 astrometric time series for the nearby {\it Gaia} LSP sample and recomputed the resulting single-star-fit {\tt RUWE}. We considered one-sided dust-obscuration models consistent with the observed {\it Gaia} $G$-band semi-amplitudes ($A_G$), and tested different time dependencies for the photocenter motion: coherent sinusoidal variability, smooth stochastic wander, and random epoch-to-epoch jitter. The detailed implementation is outlined in Appendix~\ref{app:dust_ruwe_method}.

The observed {\it Gaia} LSP semi-amplitudes in the nearby sample are $A_G=0.118^{+0.078}_{-0.061}$~mag (e.g., Appendix~\ref{app:all_plots}). In an optimistic one-sided obscuration model, these amplitudes imply maximum photocenter shifts of $0.093^{+0.088}_{-0.051}~R_\star$. Using the SED-based radii from Appendix~\ref{app:gaia_sed_fits}, this corresponds to $0.10^{+0.16}_{-0.06}$~mas for the $750$~pc sample, or about $\sim0.5^{+0.3}_{-0.4}$ times the minimum binary reflex scale inferred from the RVs (i.e., for edge-on inclinations). Thus, dust obscuration at the level implied by the observed $A_G$ values can produce photocenter shifts that are non-negligible, but they are generally smaller than those expected from binary motion.

The strongest constraints come from the nearest systems. The nearest {\it Gaia} LSP star lies at $391$~pc (Table~\ref{tab:gaia_sed_fit_results}). For this star, the one-sided dust upper limit implied by the observed $A_G$ is $0.646$~mas, whereas the corresponding random-jitter amplitude that still keeps the median predicted ${\tt RUWE}<1.1$ is only $0.07$~mas. More generally, for the full nearby {\it Gaia} LSP sample ($d<1.5$~kpc, $N=224$), the maximum random photocenter jitter compatible with median predicted ${\tt RUWE}<1.1$ is $0.08 \pm 0.02$ mas, corresponding to $\sim0.1$--$0.4~R_\star$. These thresholds are comfortably above the photocenter shifts predicted for convection in evolved cool stars \citep{Chiavassa22,Chiavassa2018}, so modest intrinsic photocenter jitter remains fully compatible with the observed {\tt RUWE} of both LSP and non-LSP pulsators.

In summary, the key result is that the astrometric impact of dust obscuration depends strongly on the time dependence of the photocenter motion. For a fixed photocenter semi-amplitude $A_G$, our {\tt gaiamock} injections show that random epoch-to-epoch jitter inflates {\tt RUWE} much more efficiently than coherent sinusoidal motion on the LSP period, while smooth or coherent variability can often be partially absorbed by the single-star astrometric fit (Appendix~\ref{app:dust_ruwe_method}). In the maximally one-sided obscuration model, $23$ of the $45$ radius-calibrated nearby stars would be expected to exceed the ${\tt RUWE}=1.1$ threshold, yet those stars have a median observed ${\tt RUWE}=1.076$, and only $26\%$ exceed ${\tt RUWE}=1.1$. We therefore conclude that the {\it Gaia} data \emph{disfavor} simple obscuration models that produce large random, non-axisymmetric photocenter excursions. However, they do not by themselves rule out other dust geometries or coherent photocenter displacements caused by obscuration on the observed amplitude $A_G$. Note also that purely radial pulsations would not shift the photocenter at all and are fully consistent with the low observed {\tt RUWE} values. {\it Gaia} {\tt RUWE} thus provides a useful astrometric benchmark for future dust-obscuration and pulsation models of LSPs, even if it is not by itself decisive for every geometry or time dependence.

\subsection{Implications for binary models}\label{subsec:implications_for_binarity}

Binary companions have long been considered one of the leading explanations for long secondary periods. In these models, the LSP is identified with the orbital period of a low-mass stellar or substellar companion, while the photometric variability is attributed to periodic obscuration by dust, modulation of circumstellar material near the companion, or tidally induced variability \citep[e.g.,][]{Wood99,Wood04,Soszynski07,Soszynski14,Soszynski21,Percy23,Goldberg24,Decin25}. 

Interpreting the observed RV variations as orbital motion implies companions with masses clustered near the stellar--substellar boundary (Figure \ref{fig:rv_amps_m2s}) on au-scale orbits (Section~\ref{subsec:results_rvamps}), in the same region as the brown dwarf desert around solar-type stars, the progenitors of most LSPs.
Reconciling this demographic with the high incidence of LSPs ($\sim30\%$) among evolved red giants has long posed a challenge for binary-based LSP models. One proposed reconciliation is that many LSP companions were initially massive planets that accreted material up to their present mass \citep[e.g.,][]{Retter05, Soszynski21}.

In addition, Keplerian fits to LSP radial-velocity curves yield a non-uniform distribution of the argument of periastron ($\omega$), which preferentially clusters at $\omega > 180^\circ$ \citep[e.g.,][]{Wood04,Nicholls09}. This implies that the red giant is usually closest to the observer at periastron while the lower-mass companion is farther away -- a configuration that is statistically unlikely for randomly oriented binary systems. Recently, \citet{Decin25} proposed that eccentric low-mass companions embedded in dusty environments, combined with time-in-dust selection effects, may reproduce the observed phase relations and periastron distribution. However, this interpretation still faces the challenge of the brown dwarf desert.

Irrespective of the hypothetical companion's origin, all companion-based LSP models share a fundamental and testable prediction: a companion massive enough to generate the observed RV amplitudes (a few ${\rm km~s^{-1}}$) is present today. For the nearby LSP population, our results show that such companions should generally induce {\it Gaia}-detectable astrometric reflex motion of the red giant, and that this motion is absent in the {\it Gaia} data. This conclusion is largely independent of the mechanism producing the photometric variability, and is robust to the companion's origin, orbital configuration, and the eccentric binary models considered in Appendix~\ref{app:varying_model}.

Dust obscuration can perturb the optical photocenter, but in the simple one-sided models we considered, it does not explain the missing binary astrometric signal in the nearby sample (Section \ref{subsec:dust_discussion} and Appendix~\ref{app:dust_ruwe_method}). If the dust-induced photocenter shifts behaved like random jitter, they would inflate {\tt RUWE}, which is not observed. Coherent or smoothly varying photocenter motion is not expected to significantly enhance {\tt RUWE}, so it is more difficult to strictly rule out. 
Another possibility is that intrinsic stellar variability could induce astrometric jitter, possibly elevating {\tt RUWE}. 
Three-dimensional radiative-hydrodynamic simulations of red supergiants predict that large convective cells can shift the stellar photocenter by $\sim1$--$5\%$ of the stellar radius, potentially contributing to {\it Gaia} astrometric errors \citep{Chiavassa22}. 
However, a detailed analysis of {\it Gaia} astrometry shows that such photocenter motions are presently undetectable: the measured excess astrometric noise in {\it Gaia} DR3 is dominated by other noise sources and is typically an order of magnitude larger than the predicted convective signal \citep{Kochanek23}. 
The absence of elevated {\tt RUWE} in LSP stars therefore does not conflict with models invoking convection or radial pulsations, although it may still disfavor variability mechanisms that produce large non-axisymmetric surface brightness structures capable of generating detectable photocenter shifts.

Taken together, our findings rule out low-mass stellar and substellar companions as the {\it dominant} cause of LSPs in giant branch stars. 
While binarity may still influence the circumstellar environment in individual systems, companion-induced orbital motion cannot account for the LSP phenomenon as a population-wide explanation, motivating renewed focus on alternative stellar mechanisms, possibly intrinsic to the star, as the origin of LSPs.

\subsection{Betelgeuse}\label{subsec:betelgeuse}

$\alpha$~Orionis (Betelgeuse), the tenth brightest star in the sky and the nearest red supergiant \citep{Hoffleit82}, is frequently discussed as an LSP system. 
Although Betelgeuse is more massive and luminous than the RGB and AGB stars that dominate our sample, it is nonetheless considered an LSP star: it shows variability on timescales of $\sim400$~days, commonly attributed to pulsation, and $\sim2100$~days, widely identified as its LSP \citep[e.g.,][]{Joyce20}. 
Betelgeuse's LSP behavior, both in photometry and RVs, is broadly similar to those of lower-mass LSP variables, making it a natural benchmark for proposed binary models.

Recent studies have interpreted Betelgeuse’s LSP as evidence for a low-mass companion \citep{Goldberg24,MacLeod25}. 
Using century-long RV and astrometric data, \citet{MacLeod25} propose a companion of mass $\lesssim1~{\rm M_\odot}$ on a $\sim2110$~day orbit at a separation of $\sim2$--$3\times$ Betelgeuse’s radius. 
Although this companion mass is larger than the $\sim0.1~{\rm M_\odot}$ companions implied for typical LSP stars in our sample, the inferred mass ratio ($M_2/M_1 \sim 0.05$--$0.1$) and measured RV semi-amplitude ($K \approx 1.5 \pm 0.34~{\rm km~s^{-1}}$) lie squarely at the peak of the LSP RV amplitude distribution (Figure \ref{fig:rv_amps_m2s}).
Their model invokes tidal spin--orbit coupling and a dust-rich wake trailing the companion to explain Betelgeuse’s anomalously rapid rotation and photometric phase offset, similar to earlier interpretations based on circumstellar dust modulation \citep[e.g.,][]{Goldberg24}. However, as \citet{MacLeod25} discuss, the astrometric signal could also be modeled as large-amplitude white noise, and more data will be needed to test the binary interpretation. \citet{Goldberg24} measure similar companion properties of $1.17 \pm 0.07~{\rm M_\odot}$ at a radius of  $2.43^{+0.21}_{-0.32}$ times the radius of Betelgeuse\footnote{Antares, another one of the closest supergiants \citep[$d=170$~pc;][]{vanLeeuwen07}, also shows an LSP ($P \simeq 2167$~days) with an $K\approx2.73~{\rm km~s^{-1}}$, implying a low-mass companion ($M_2 \sin i/M_1 \approx 0.07$) if interpreted as orbital motion \citep{Pugh13}.}. 

Optical speckle observations of Betelgeuse obtained near the predicted quadrature reported a probable point source at the expected separation ($\approx52$~mas) and position angle, although at alarmingly low significance ($\sim1.5\sigma$; \citealt{Howell25}). 
Additional spectroscopic work has interpreted time-variable absorption and outflow signatures as evidence for a companion-induced wake within the extended atmosphere of the star \citep{Dupree26}. Recent direct searches have also placed constraints on the proposed companion. 
Deep {\it Chandra} observations obtained near the predicted orbital quadrature detected no X-ray source at the position of Betelgeuse, implying an upper limit of $L_X \lesssim 2\times10^{30}$ erg s$^{-1}$ and ruling out an accreting compact object such as a white dwarf or neutron star \citep{OGrady25}. 
Complementary far-UV spectroscopy with {\it HST}/STIS likewise detected no spectral features at the predicted velocity of the companion and excludes companions with masses $\gtrsim1.5\,M_\odot$ or FUV emission exceeding $\sim10^{-14}$ erg s$^{-1}$ cm$^{-2}$ \AA$^{-1}$ in the $1200$--$1700$ \AA\ band \citep{Goldberg25}. 
These observations, therefore, rule out several classes of luminous or accreting companions but remain consistent with a faint low-mass young stellar object.
Alternative interpretations attribute Betelgeuse’s long period to intrinsic stellar variability. Non-adiabatic pulsation models suggest that the $\sim2100$~day signal may correspond to the fundamental radial mode of the star rather than to orbital motion \citep{Saio23}.

Our results provide a population-level test of the binary interpretation for LSPs, which {\it disfavors} low-mass companions as the dominant cause of LSPs in evolved giants. Given that Betelgeuse’s RV amplitude, photometric amplitude, period, and implied mass ratio are typical of the LSP population, it seems plausible that its LSP may likewise arise from an alternative stellar variability mechanism rather than from orbital motion due to a binary companion. 
However, Betelgeuse and LSP supergiants alike are substantially more massive and luminous than the stars studied in our sample, so they could be treated as a different regime.
Betelgeuse may also represent a particular case where binarity does cause the LSP.
Continued monitoring, particularly near future orbital quadratures, will therefore be crucial for determining whether Betelgeuse's LSP is an exceptional case of binarity or whether its variability reflects the same physical processes operating in the broader LSP population.

\section{Conclusions}\label{sec:conclusions}

More than a century after their discovery, the origin of long secondary periods (LSPs) remains unknown, despite being observed in $\sim1/3$ of pulsating luminous RGB and AGB stars. 
Low-mass stellar or substellar companions have emerged as one of the leading explanations for LSPs, particularly in recent years \citep[e.g.,][]{Soszynski21,Goldberg24,MacLeod25,Decin25}. In these models, the LSP corresponds to a companion's orbital period, while the photometric variability is attributed to occultation or modulation of circumstellar material. In this work, we provide a direct, population-wide test of the binary LSP hypothesis using a combination of RVs, photometry, and astrometry available from the {\it Gaia} Focused Product Release \citep{Gaia_FPR}.

Using RVs for a sample of nearby LSP stars (e.g., Figure \ref{fig:example_rv_phot}), we infer the companion masses ($M_2$) implied under the binary hypothesis. The observed RV semi-amplitudes cluster sharply around $K\approx1.6~{\rm km~s^{-1}}$, corresponding to inferred companion masses narrowly peaked near $M_2\approx0.09~{\rm M_\odot}$ (Figure \ref{fig:rv_amps_m2s}). We then forward-model the expected astrometric signatures of these hypothetical companions using {\tt gaiamock} \citep{ElBadry24} and compare them to the observed {\it Gaia} astrometry. We find that the predicted re-normalized unit weight error ({\tt RUWE}) values are systematically larger than those observed (Figure~\ref{fig:RUWE}). For $80\%$ of the local sample ($d<750$~pc), the expected astrometric signal from binarity is {\it not} present in observations. In contrast, known ellipsoidal-variable binaries show elevated {\tt RUWE} in {\it Gaia}, consistent with forward-model predictions based on their RV amplitudes.

We conclude that low-mass stellar or substellar companions are {\it not} the dominant cause of LSPs in evolved giant branch stars. 
The result applies to all companions in the RV-inferred mass range, irrespective of their specific variability-causing mechanism, orbital configuration, or origin. Namely, invoking different eccentricities, inclinations, and modulation of circumstellar dust over reasonable ranges does not change our conclusions (Appendix \ref{app:varying_model} and Section \ref{subsec:dust_discussion}). 

Some of the nearest and brightest red supergiants, such as Betelgeuse, show long-term variability attributed to LSPs \citep[e.g.,][]{Kiss06,Joyce20,Goldberg24, MacLeod25}.
Given that Betelgeuse shows an RV amplitude, photometric amplitude, period, and implied mass ratio typical of the broader LSP population \citep[e.g.,][]{Goldberg24,MacLeod25}, our results suggest that its LSP may likewise arise from an alternative physical mechanism rather than from orbital motion due to a binary companion, although it may represent a special case.

Looking ahead, the empirical constraints established by the {\it Gaia} sample and many previous works provide an extensive list of LSP properties -- such as RV amplitudes, phase offsets between RV and photometric variability, IR behavior, and now astrometric behavior -- that successful LSP models should reproduce. 
Our astrometric constraints motivate a renewed focus on mechanisms that do not require a low-mass binary companion as the driver of LSPs. 

\section{Acknowledgments}\label{acknowledgments}
C.S. acknowledges support from the Department of Energy Computational Science Graduate Fellowship.  This
research was supported in part by grant NSF PHY2309135 to the Kavli Institute for Theoretical Physics
(KITP). This research was supported by NSF grants AST-2540180 and AST-2307232.
This material is based upon work supported by the U.S. Department of Energy, Office of Science, Office of Advanced Scientific Computing Research, under Award Number DE-SC0026073. M.M. is grateful to A. Dupree, S. Blunt, and A. Khwaja for insight and discussions about the nature of LSPs.
This work made use of \texttt{\href{https://github.com/cheyanneshariat/OverCite}{OverCite}} \citep{Shariat_OverCite_2026}, an in-editor citation tool for \LaTeX.

\appendix 
\twocolumngrid

\section{SED fitting}\label{app:gaia_sed_fits}

We estimate the stellar radii and effective temperatures for the nearby {\it Gaia} LSP sample by fitting simple two-parameter photospheric SED models to the {\it Gaia} DR3 $G$, $G_{\rm BP}$, and $G_{\rm RP}$ photometry.  The SEDs were modeled with single-star BaSeL atmospheres \citep{Lejeune1997,Lejeune1998,Westera2002} using the \texttt{pystellibs} package\footnote{\url{https://github.com/mfouesneau/pystellibs}}, and synthetic {\it Gaia} photometry was computed with \texttt{pyphot}. We fit two free parameters, $T_{\rm eff}$ and $R_\star$. The stellar mass was fixed to $1.2~{\rm M_\odot}$. For each source, we fixed the line-of-sight extinction to the value from the Edenhofer 3D dust map \citep{Edenhofer24}, expressed as $A_V = 3.1\,E(B-V)_{\rm edenhofer}$, and we fixed the distance to the value from its {\it Gaia} parallax. The adopted priors were uniform over $2000 \le T_{\rm eff}/{\rm K} \le 4500$ and $5 \le R_\star/R_\odot \le 350$. We sampled the posterior with \texttt{emcee} \citep{emcee} using $96$ walkers, $500$ burn-in steps, and $1400$ production steps per star, and adopted the posterior medians and 16th/84th percentiles as the fitted values and formal $1\sigma$ uncertainties. Note that reported uncertainties should be regarded as formal statistical errors only. They do not include systematic uncertainties in the distances, extinction values, possible atmosphere-model systematics, or the limitations of fitting only three broad {\it Gaia} bands. The quoted $T_{\rm eff}$ and $R_\star$ uncertainties are therefore underestimated, but are sufficient for our limited purpose of expressing the astrometric constraints in units of $R_\star$ in Section~\ref{subsec:dust_discussion}.

\begin{figure}
\centering
\includegraphics[width=0.99\columnwidth]{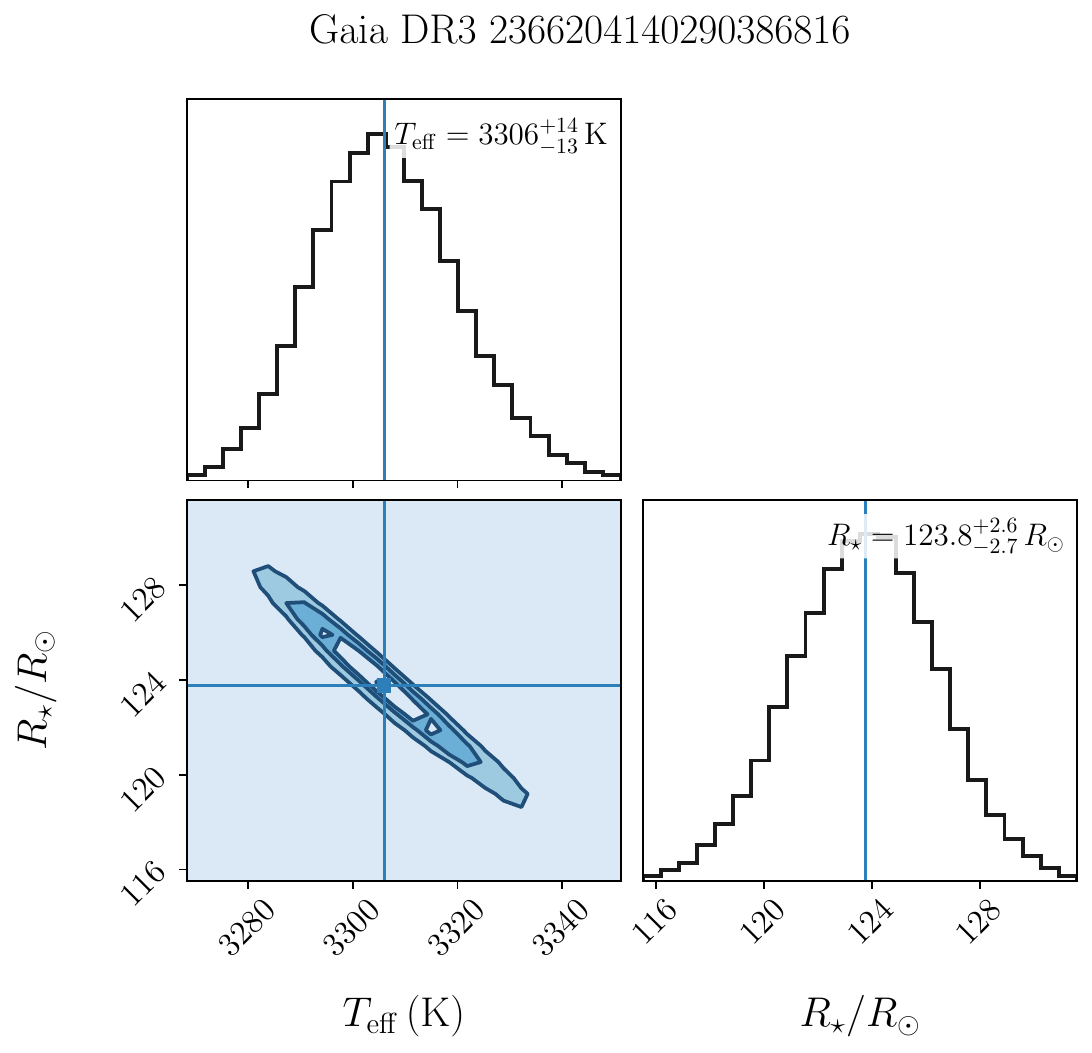}
\caption{Representative posterior corner plot for one {\it Gaia} BP/G/RP SED fit. The fitted parameters are $T_{\rm eff}$ and $R_\star$, with the extinction fixed to the dust map value and the stellar mass fixed to $1.2~{\rm M_\odot}$. The blue lines mark the posterior medians.}
\label{fig:gaia_sed_fit_examples}
\end{figure}

Figure~\ref{fig:gaia_sed_fit_examples} shows a representative posterior corner plot, and Table~\ref{tab:gaia_sed_fit_results} lists the {\it Gaia} DR3 source id, distance, fixed Edenhofer extinction, and posterior median $T_{\rm eff}$ and $R_\star$ values for all $45$ fitted stars. A machine-readable version of this table is provided online\footnote{\texttt{gaia\_lsp\_750pc\_sed\_fit\_machine\_readable.csv} at \url{https://zenodo.org/records/19412352}}.

\section{Varying binary model assumptions}\label{app:varying_model}

Our fiducial {\tt gaiamock} forward model in Section~\ref{subsec:results_ruwe} assumes circular orbits. Here we test two alternative assumptions motivated by recent binary models for LSPs: eccentric orbits and viewing geometries biased toward edge-on configurations. Specifically, we repeat the DR3 {\tt RUWE} forward modeling for fixed $e=0.3$, for $e\sim U(0.1,0.6)$, and for the same eccentricity prescriptions combined with an edge-on inclination prior ($45^\circ < i < 135^\circ$). In all cases, the companion masses are derived using the observed RV amplitudes and Equation \eqref{eq:rv_mass_func}. We simulate $50$ realizations to generate $1\sigma$ errors, sampling eccentricities, inclinations, and orbital angles each time.

\begin{figure*}[ht!]
\centering
\includegraphics[width=0.49\textwidth]{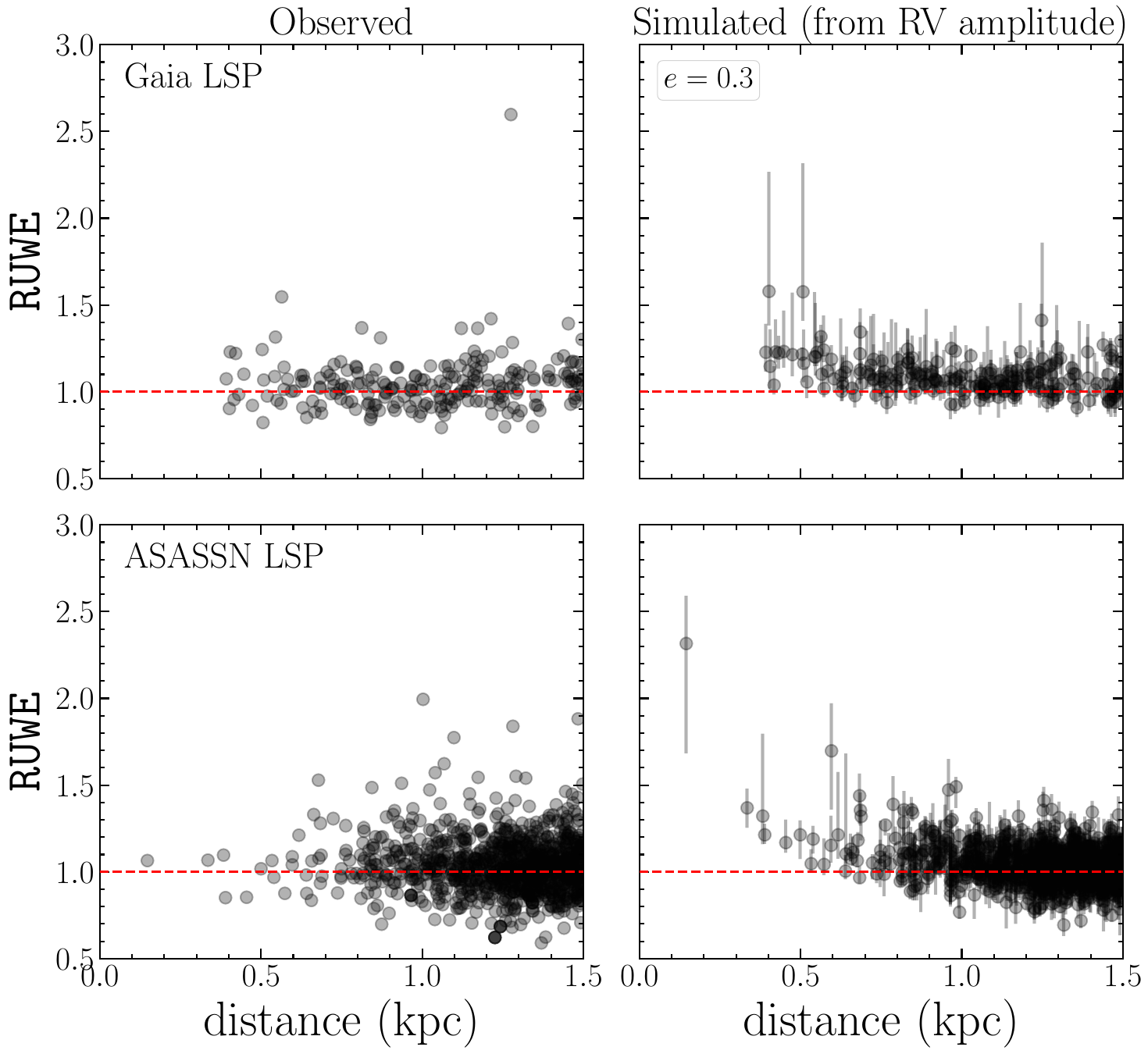}
\includegraphics[width=0.49\textwidth]{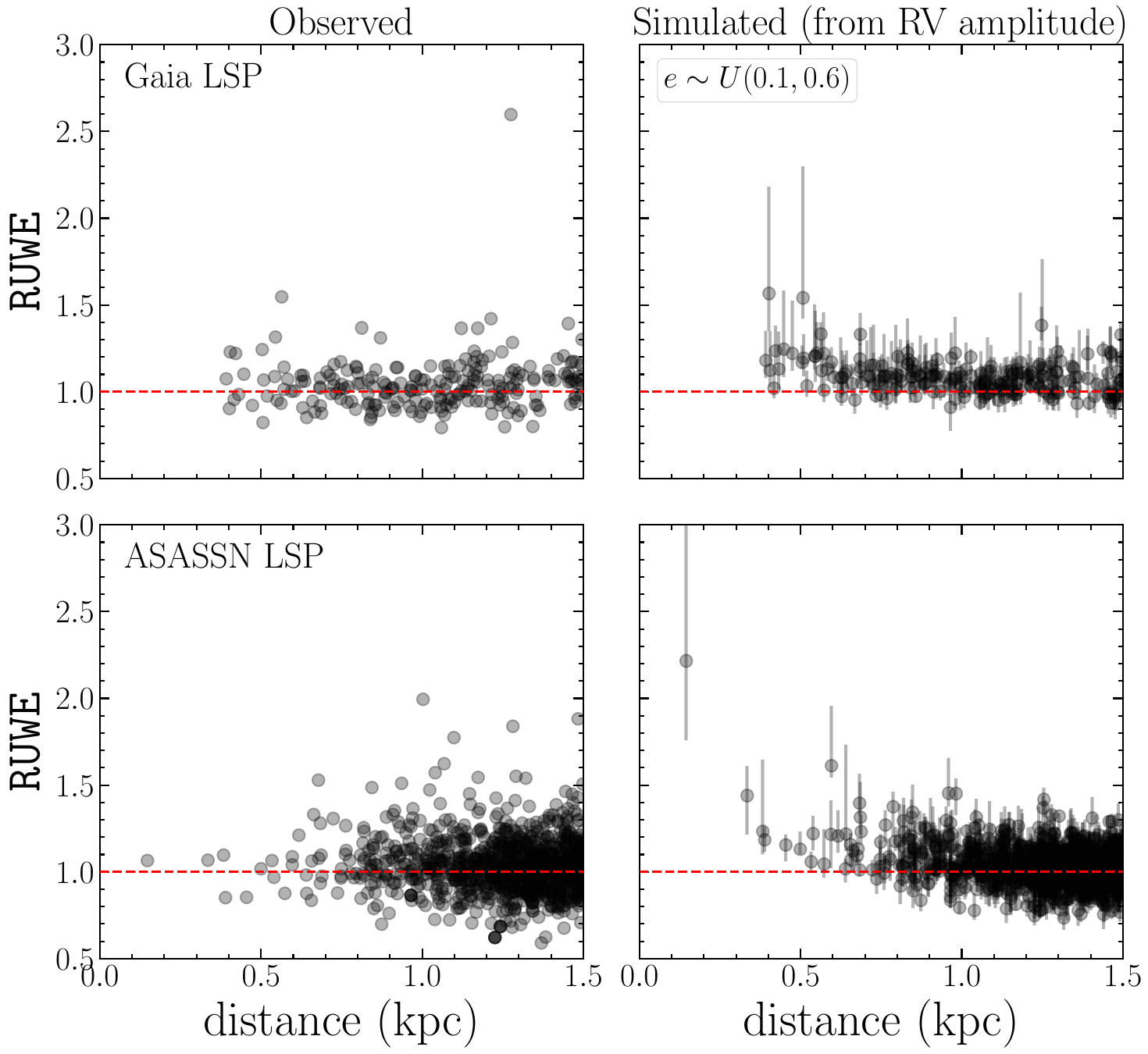}\\
\includegraphics[width=0.49\textwidth]{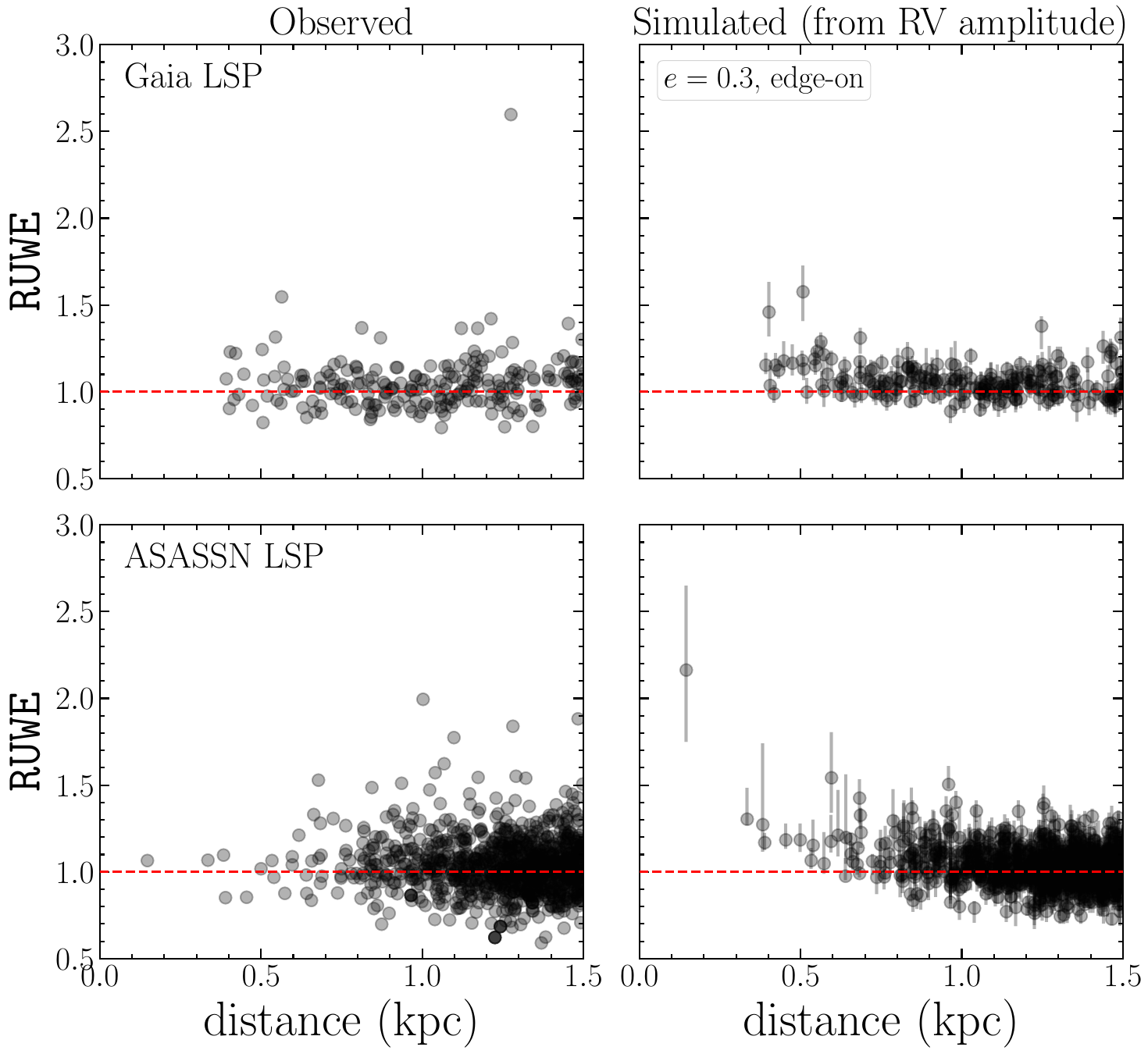}
\includegraphics[width=0.49\textwidth]{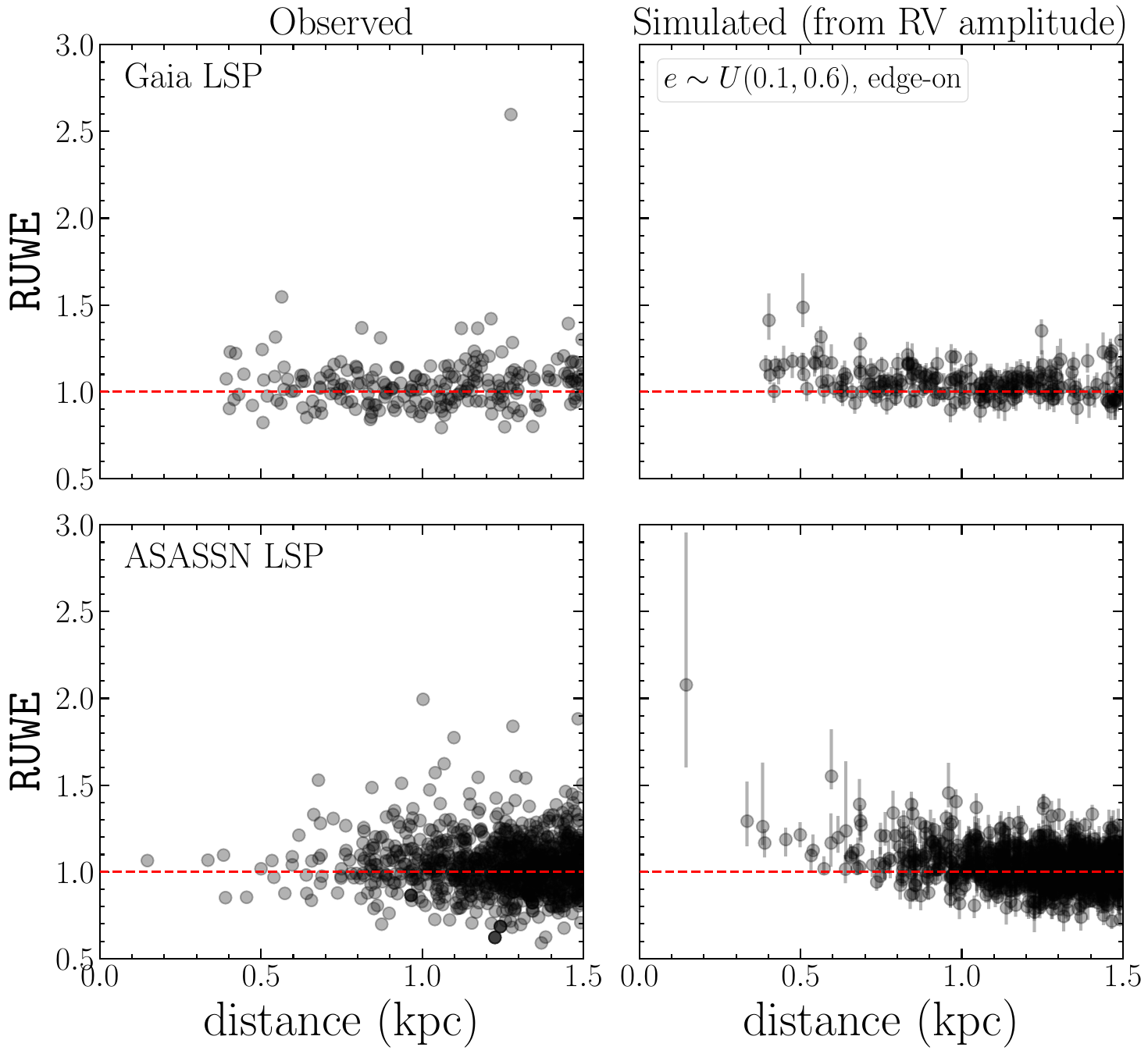}
\caption{RUWE predictions for alternative binary-model assumptions. The top row shows DR3 forward models for eccentric binaries with isotropic inclinations, assuming either fixed $e=0.3$ (left) or $e\sim U(0.1,0.6)$ (right). The bottom row shows the corresponding DR3 predictions when the same eccentricity prescriptions are combined with an edge-on inclination prior, $45^\circ < i < 135^\circ$. In each panel, the left-hand column shows the observed {\it Gaia} {\tt RUWE} for the {\it Gaia} and ASAS-SN LSP samples, and the right-hand column shows the {\tt RUWE} predicted under the binary hypothesis for the stated assumptions. Changing the eccentricity or inclination distribution shifts the predicted {\tt RUWE} modestly, but nearby LSPs are still generally expected to have larger {\tt RUWE} than are observed.}\label{fig:ruwe_variations}
\end{figure*}

Figure~\ref{fig:ruwe_variations} shows the results of the simulations. We find that including eccentricity does not change the overall conclusions: nearby LSPs are still predicted to have elevated {\tt RUWE} values. Furthermore, restricting the sample to preferentially edge-on inclinations reduces the {\tt RUWE} uncertainties, strengthening the statistical significance of the discrepancy with observed LSP {\tt RUWE} values.
Thus, our main conclusion is unchanged: varying the binary model assumptions within the range proposed in the recent literature does not make the low-mass-companion interpretation consistent with the astrometry.

\section{Gaia DR4 predictions}\label{app:gaia_dr4}

\begin{figure}
\centering
\includegraphics[width=0.99\columnwidth]{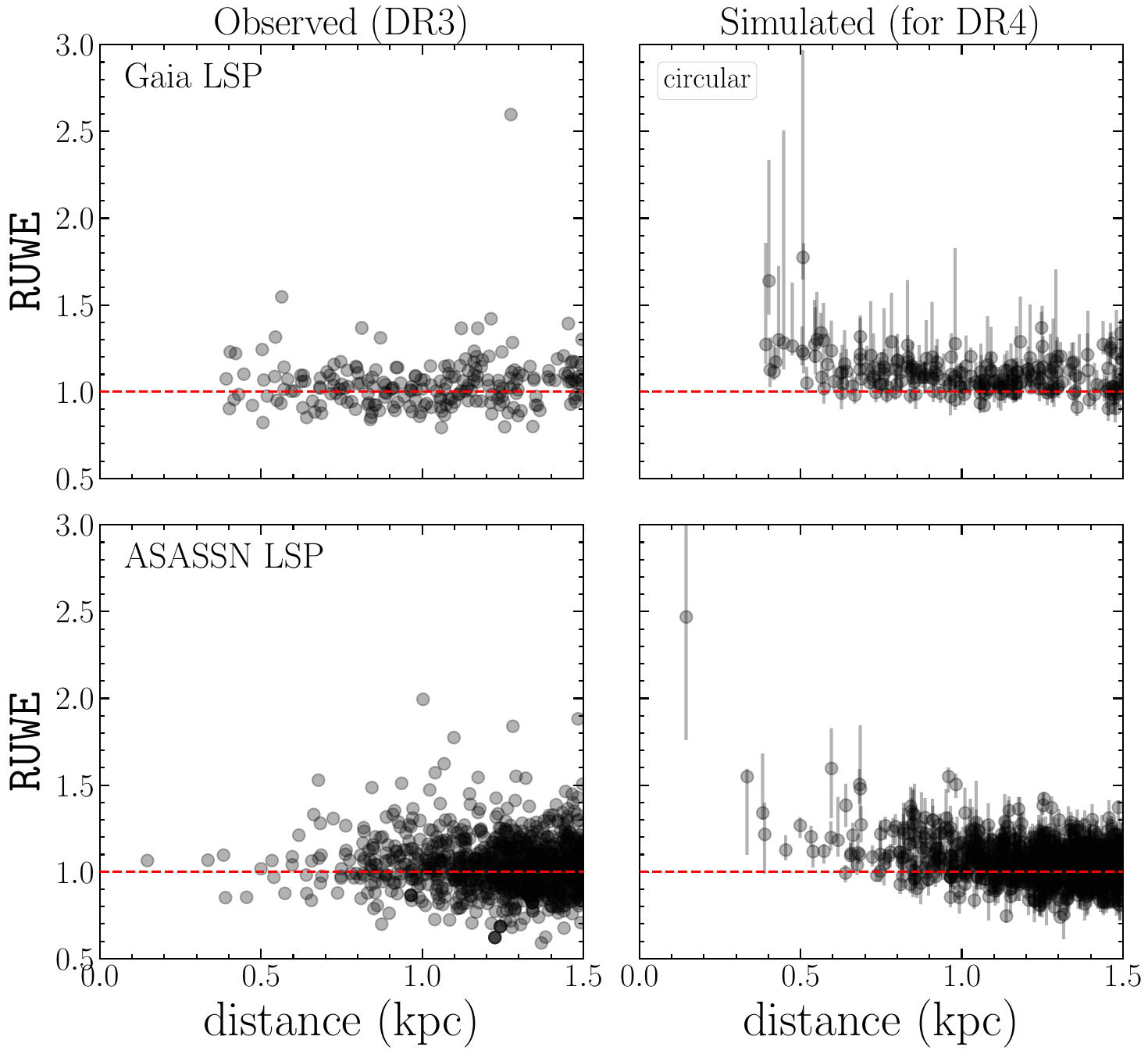}
\caption{{\tt RUWE} for LSP stars assuming the binary hypothesis in {\it Gaia} DR3 (left) and DR4 (right). The binary assumptions are the same as in Figure \ref{fig:RUWE}.}\label{fig:ruwe_all_dr4}
\end{figure}

Figure~\ref{fig:ruwe_all_dr4} shows the expected {\it Gaia} DR4 analogue of our fiducial RUWE test. In the right-hand panels, we rerun the same forward model used in the main text, but with {\tt gaiamock} configured for DR4 rather than DR3. The underlying binary assumptions are unchanged: the orbital period is fixed to the observed LSP, the companion properties are inferred from the RV variability, and the predicted astrometric residuals are converted to the single-star-fit goodness-of-fit statistic reported as {\tt RUWE}. The left-hand panels show the currently available observed DR3 {\tt RUWE} values for the same {\it Gaia} and ASAS--SN LSP samples. Overall, {\it Gaia} DR4 will not provide a significant change to the predicted {\tt RUWE} values. However, {\it Gaia} DR4 will decrease the scatter in {\tt RUWE} for single stars, leading to an increase in the discrepancy between observed and simulated LSP {\tt RUWE} values. 

\section{Dust-obscuration photocenter shifts}\label{app:dust_ruwe_method}

In Section~\ref{subsec:dust_discussion}, we test whether dust obscuration or intrinsic stellar variability could perturb the optical photocenter strongly enough to affect the {\it Gaia} DR3 astrometry of nearby LSP stars. Here we outline how this was implemented.

For each source, we first generate a baseline synthetic single-star astrometric time series with {\tt gaiamock}, using the observed sky position, parallax, proper motion, and $G$-band magnitude. This yields the astrometric epochs $t_j$, scan angles $\psi_j$, parallax factors, synthetic along-scan measurements $\eta_j$, and corresponding along-scan uncertainties $\sigma_{\eta,j}$ expected for a single star in {\it Gaia} DR3. We then inject a prescribed photocenter offset into the synthetic astrometry and recompute the resulting single-star-fit {\tt RUWE}. No binary orbital motion is included in these tests.

To connect the astrometric perturbation to the observed photometric variability, we adopt a simple one-sided dust-obscuration model. The stellar disk is treated as a uniform circle of radius $R_\star$, and the hemisphere with $x>0$ is attenuated by a factor $(1-\delta)$ while the opposite side is unchanged. For an unobscured flux $F_0$, the obscured flux is therefore
\begin{equation}
F = \frac{F_0}{2} + (1-\delta)\frac{F_0}{2}
= F_0\left(1-\frac{\delta}{2}\right).
\end{equation}
The corresponding magnitude change is
\begin{equation}
\Delta m = -2.5\log_{10}\left(\frac{F}{F_0}\right)
= -2.5\log_{10}\left(1-\frac{\delta}{2}\right).
\end{equation}
Because the {\it Gaia} catalog reports the $G$-band semi-amplitude ($A_G$), we identify
\begin{equation}
A_G = \frac{\Delta m}{2}
= -1.25\log_{10}\left(1-\frac{\delta}{2}\right).
\end{equation}

For each value of $\delta$, we compute the corresponding photocenter displacement from the first moment of the attenuated surface-brightness distribution,
\begin{equation}
x_{\rm ph} =
\frac{\iint x\,I(x,y)\,dx\,dy}{\iint I(x,y)\,dx\,dy},
\end{equation}
which defines a maximum one-sided photocenter shift in units of the stellar radius,
$
\Delta x_{\rm ph}/R_\star = f(A_G).
$
We evaluate $f(A_G)$ numerically.
To express the same displacement in angular units, we use the stellar radii estimated from the SED fits described in Appendix~\ref{app:gaia_sed_fits}. For a star of radius $R_\star$ at distance $d$, the angular stellar radius is
\begin{equation}
\theta_\star({\rm mas}) = 4.65047\,\left(\frac{R_\star}{R_\odot}\right)
\left(\frac{d}{{\rm pc}}\right)^{-1},
\end{equation}
so the corresponding angular photocenter shift is
\begin{equation}
\Delta x_{\rm ph}({\rm mas}) =
\left(\frac{\Delta x_{\rm ph}}{R_\star}\right)\theta_\star.
\end{equation}
We then assign a time dependence to the photocenter motion. At each astrometric epoch $t_j$, the sky-plane photocenter offset in R.A. and Dec. is written as $(\Delta\alpha_j,\Delta\delta_j)$, in mas. 

\begin{figure*}
\centering
\includegraphics[width=0.98\textwidth]{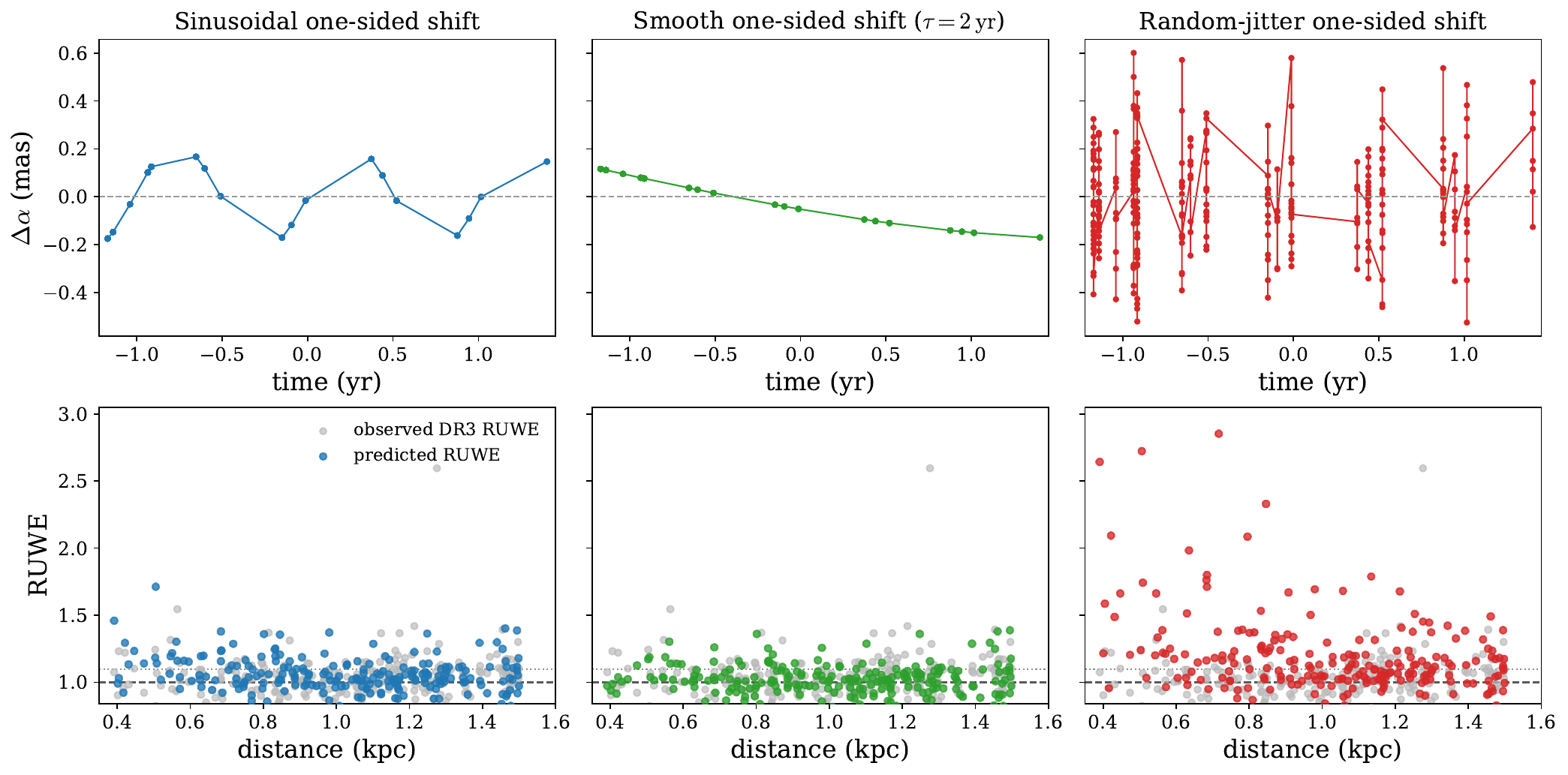}
\caption{Dust-photocenter time models and their predicted astrometric signatures. The top row shows example photocenter motion in $\Delta\alpha$ for one representative LSP star, using the same dust-inferred photocenter semi-amplitude but different time dependences: sinusoidal motion on the LSP timescale (left), smooth stochastic wander with coherence time $\tau=2$ yr (middle), and random epoch-to-epoch jitter (right). The bottom row shows the corresponding predicted {\tt RUWE} values as a function of distance for the full {\it Gaia} LSP sample within $1.5$~kpc, together with the observed DR3 {\tt RUWE} values. For a fixed photocenter semi-amplitude, random jitter inflates {\tt RUWE} more efficiently than smooth or sinusoidal motion.}
\label{fig:dust_time_models_ruwe}
\end{figure*}

We test three cases of dust variability: sinusoidal motion, smooth stochastic wander, and random jitter (top row of Figure \ref{fig:dust_time_models_ruwe}). In the sinusoidal case, the photocenter varies coherently on the LSP timescale:
\begin{equation}
\Delta\alpha_j = A_{\rm ph}\sin\left[\frac{2\pi (t_j-t_0)}{P_{\rm LSP}}+\phi_0\right], \qquad
\Delta\delta_j = 0,
\end{equation}
where $A_{\rm ph}=\Delta x_{\rm ph}({\rm mas})$. In the random-jitter case, we draw independent Gaussian values for both coordinates and rescale the realization so that
\begin{equation}
\sqrt{\left\langle \Delta\alpha^2 + \Delta\delta^2 \right\rangle} = A_{\rm ph}.
\end{equation}
For the smooth-wander case, we begin from Gaussian white noise in both coordinates, convolve each time series with a Gaussian kernel in time, and then rescale the result to the same RMS amplitude $A_{\rm ph}$. The smooth-wander and random-jitter models, therefore, differ only in temporal coherence, while the sinusoidal case represents the limiting case of fully coherent variability on the LSP timescale.
The top panel of Figure~\ref{fig:dust_time_models_ruwe} shows the representative photocenter shifts over time for the different obscuration models.

Because {\it Gaia} measures the along-scan coordinate rather than $(\Delta\alpha,\Delta\delta)$ directly, we project the sky-plane photocenter offset onto the scan direction at each epoch:
\begin{equation}
\Delta\eta_j = -\Delta\delta_j\cos\psi_j - \Delta\alpha_j\sin\psi_j.
\end{equation}
We then perturb the synthetic along-scan astrometry according to
\begin{equation}
\eta_j^\prime = \eta_j + \Delta\eta_j,
\end{equation}
and pass the modified time series back through the same single-star astrometric fit used by {\tt gaiamock} to compute the resulting {\tt RUWE}. Thus, the full calculation is

\begin{equation}
\begin{aligned}
A_G &\rightarrow \delta \rightarrow \frac{\Delta x_{\rm ph}}{R_\star}
\rightarrow \Delta x_{\rm ph}({\rm mas}) \\
&\rightarrow (\Delta\alpha_j,\Delta\delta_j)
\rightarrow \Delta\eta_j \rightarrow {\tt RUWE}.
\end{aligned}
\end{equation}

The bottom row of Figure \ref{fig:dust_time_models_ruwe} shows that the astrometric response depends on both the amplitude and time dependence of the photocenter motion. Smooth or sinusoidal variability is generally difficult to distinguish beyond the closest LSPs, and can be partially absorbed by the single-star astrometric fit. On the other hand, random epoch-to-epoch jitter leaves larger residuals and therefore inflates {\tt RUWE} more efficiently for the same underlying photocenter amplitude, even out to $d\sim1.5$~kpc.

\input{tables/gaia_lsp_750pc_sed_fit_table.tex}

\section{All RV and photometric time series}\label{app:all_plots}
In Figures \ref{fig:all_phases_1} and \ref{fig:all_phases_2} we show the phase-folded RV and light curves for all targets in the $750$~pc sample.

\begin{figure*}
\centering
\includegraphics[width=0.8\textwidth]{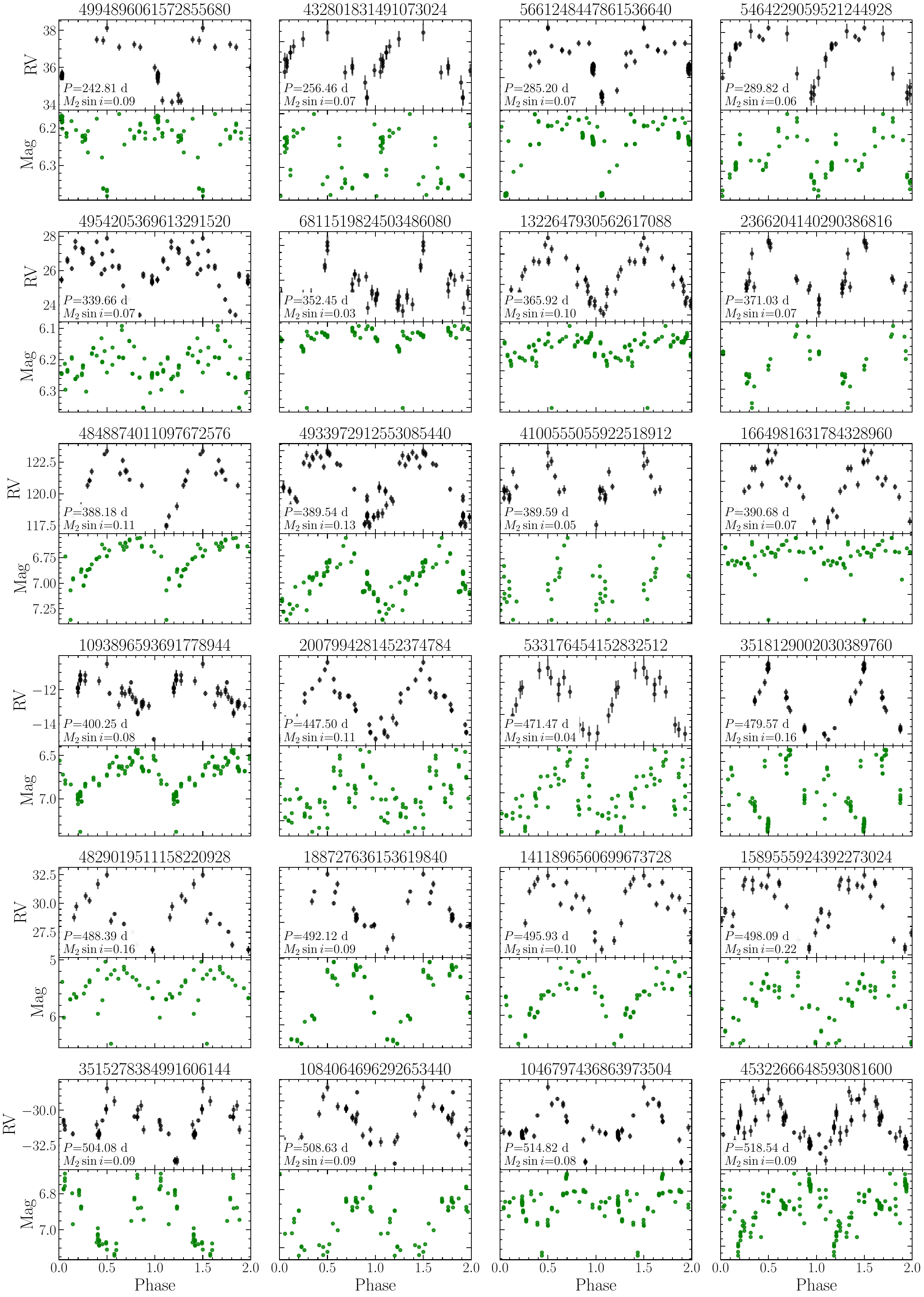}
\caption{Phase-folded RV and light curves for all {\it Gaia} LSPs within $750$~pc.
}
\label{fig:all_phases_1}
\end{figure*}

\begin{figure*}
\centering
\includegraphics[width=0.8\textwidth]{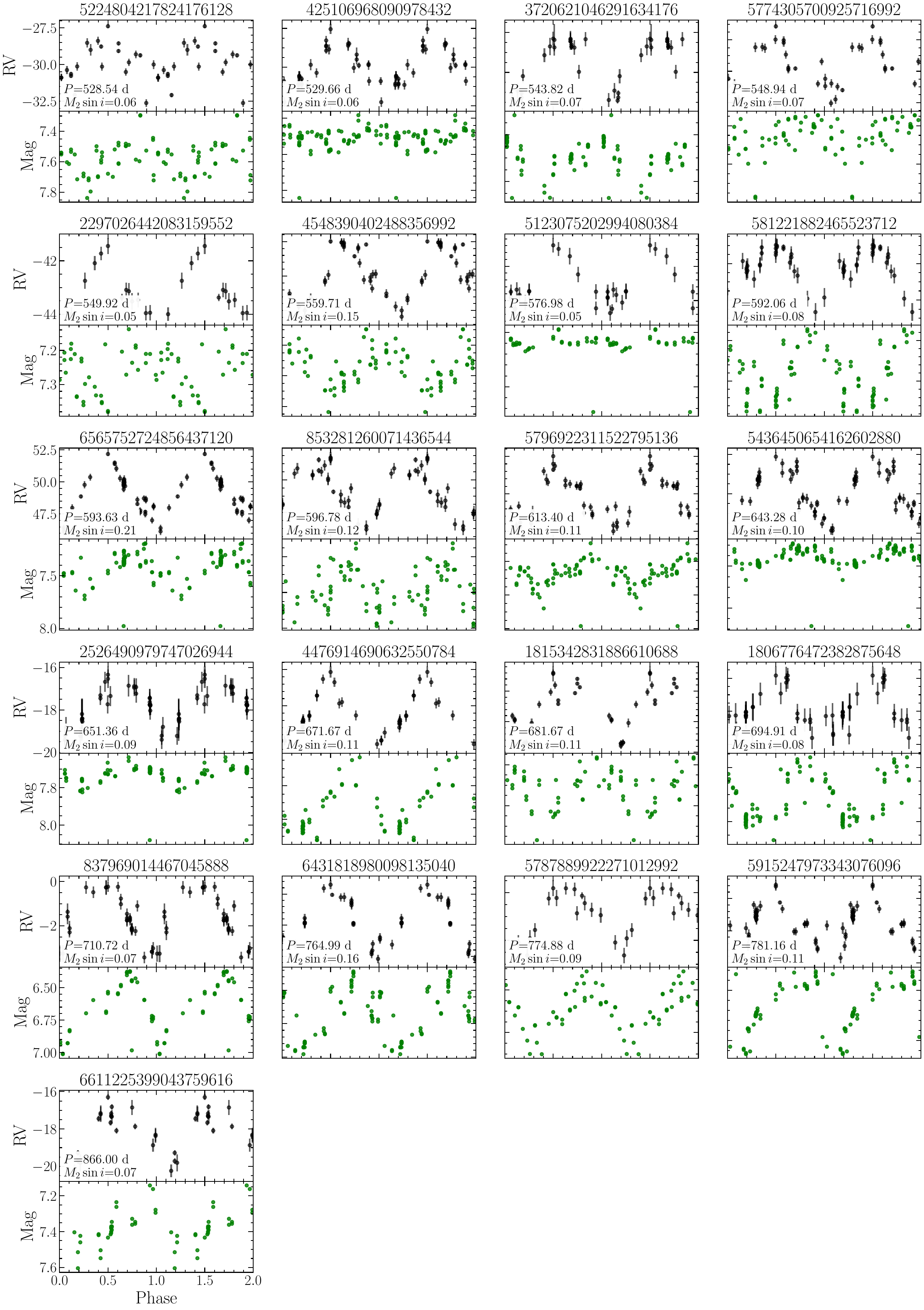}
\caption{{\bf Continuation of Figure \ref{fig:all_phases_1}.}
}
\label{fig:all_phases_2}
\end{figure*}

\clearpage
\bibliography{references}

\end{document}

%% file: tables/gaia_lsp_750pc_sed_fit_table.tex
\startlongtable
\begin{deluxetable*}{rcccc}
\tabletypesize{\scriptsize}
\tablecaption{Gaia SED-fit results for the nearby ($d<750$~pc) LSP sample.\label{tab:gaia_sed_fit_results}}
\tablewidth{0pt}
\tablehead{\colhead{Gaia DR3 source id} & \colhead{$d$ (pc)} & \colhead{$A_V$ (mag)} & \colhead{$T_{\rm eff}$ (K)} & \colhead{$R_\star\ (R_\odot)$}}
\startdata
4829019511158220928 & 391.3 & 0.041 & $3263_{-12}^{+13}$ & $205.4_{-6.1}^{+6.1}$ \\
1815342831886610688 & 400.9 & 0.058 & $3193_{-9}^{+7}$ & $146.7_{-1.5}^{+1.8}$ \\
2366204140290386816 & 404.4 & 0.059 & $3306_{-13}^{+14}$ & $123.8_{-2.7}^{+2.6}$ \\
533176454152832512 & 416.5 & 1.015 & $3459_{-11}^{+9}$ & $97.6_{-1.3}^{+1.7}$ \\
1093896593691778944 & 420.9 & 0.094 & $3159_{-12}^{+13}$ & $141.1_{-2.4}^{+2.4}$ \\
6611225399043759616 & 430.8 & 0.064 & $2846_{-24}^{+28}$ & $155.4_{-4.1}^{+3.6}$ \\
853281260071436544 & 446.4 & 0.038 & $3067_{-7}^{+7}$ & $168.9_{-1.9}^{+1.9}$ \\
3518129002030389760 & 472.9 & 0.124 & $3355_{-6}^{+5}$ & $124.0_{-1.4}^{+1.4}$ \\
4994896061572855680 & 503.3 & 0.043 & $3518_{-3}^{+3}$ & $118.6_{-0.7}^{+0.7}$ \\
6431818980098135040 & 505.3 & 0.196 & $3275_{-7}^{+7}$ & $187.3_{-3.0}^{+3.2}$ \\
3515278384991606144 & 508.2 & 0.116 & $3180_{-12}^{+11}$ & $158.6_{-2.5}^{+2.7}$ \\
6811519824503486080 & 518.3 & 0.086 & $3692_{-5}^{+5}$ & $87.3_{-0.6}^{+0.6}$ \\
4548390402488356992 & 542.5 & 0.209 & $3064_{-6}^{+6}$ & $149.9_{-1.5}^{+1.4}$ \\
4933972912553085440 & 544.8 & 0.050 & $3372_{-12}^{+14}$ & $115.1_{-2.9}^{+2.9}$ \\
4532266648593081600 & 549.7 & 0.429 & $3140_{-9}^{+9}$ & $153.4_{-1.9}^{+1.9}$ \\
5661248447861536640 & 561.0 & 0.166 & $3566_{-3}^{+3}$ & $86.0_{-0.4}^{+0.4}$ \\
1046797436863973504 & 563.7 & 0.038 & $3488_{-1}^{+1}$ & $136.6_{-0.6}^{+0.6}$ \\
6565752724856437120 & 570.6 & 0.039 & $3195_{-2}^{+2}$ & $135.8_{-0.5}^{+0.5}$ \\
4954205369613291520 & 571.1 & 0.054 & $3823_{-6}^{+6}$ & $87.2_{-0.6}^{+0.5}$ \\
1084064696292653440 & 582.3 & 0.477 & $3288_{-8}^{+8}$ & $142.7_{-1.9}^{+1.9}$ \\
3720621046291634176 & 594.4 & 0.058 & $3308_{-6}^{+7}$ & $149.2_{-1.6}^{+1.6}$ \\
2297026442083159552 & 603.1 & 0.520 & $3322_{-9}^{+10}$ & $142.0_{-2.1}^{+1.9}$ \\
1589555924392273024 & 623.7 & 0.055 & $3491_{-3}^{+3}$ & $138.4_{-1.3}^{+1.4}$ \\
5123075202994080384 & 625.9 & 0.062 & $3488_{-2}^{+2}$ & $131.2_{-0.5}^{+0.5}$ \\
5787889922271012992 & 628.6 & 0.464 & $3367_{-6}^{+6}$ & $217.8_{-3.1}^{+3.0}$ \\
1322647930562617088 & 629.9 & 0.053 & $3416_{-8}^{+10}$ & $101.7_{-1.5}^{+1.3}$ \\
188727636153619840 & 635.0 & 0.505 & $3469_{-19}^{+13}$ & $129.3_{-3.6}^{+4.8}$ \\
581221882465523712 & 641.6 & 0.079 & $3366_{-6}^{+6}$ & $116.8_{-1.4}^{+1.4}$ \\
5224804217824176128 & 659.0 & 1.044 & $3211_{-3}^{+4}$ & $195.9_{-1.9}^{+1.7}$ \\
4100555055922518912 & 666.9 & 0.737 & $3554_{-10}^{+9}$ & $90.8_{-1.1}^{+1.3}$ \\
5464229059521244928 & 682.4 & 0.158 & $3035_{-3}^{+3}$ & $247.2_{-1.7}^{+1.7}$ \\
5436450654162602880 & 682.4 & 0.167 & $3428_{-5}^{+6}$ & $116.5_{-1.1}^{+1.0}$ \\
837969014467045888 & 683.4 & 0.055 & $3352_{-26}^{+21}$ & $179.8_{-6.9}^{+3.1}$ \\
4476914690632550784 & 684.0 & 0.489 & $3214_{-3}^{+3}$ & $210.0_{-1.8}^{+1.7}$ \\
5796922311522795136 & 687.2 & 0.375 & $3164_{-6}^{+7}$ & $137.3_{-1.2}^{+1.2}$ \\
1411896560699673728 & 704.1 & 0.056 & $3473_{-15}^{+7}$ & $119.7_{-1.9}^{+2.9}$ \\
432801831491073024 & 712.7 & 0.529 & $3209_{-2}^{+2}$ & $248.1_{-1.6}^{+1.5}$ \\
5915247973343076096 & 716.4 & 0.381 & $3211_{-3}^{+3}$ & $205.7_{-1.8}^{+1.8}$ \\
5774305700925716992 & 725.2 & 0.364 & $3465_{-12}^{+10}$ & $169.7_{-3.1}^{+3.9}$ \\
425106968090978432 & 725.4 & 0.637 & $3247_{-3}^{+3}$ & $134.0_{-0.7}^{+0.7}$ \\
1806776472382875648 & 725.7 & 0.186 & $3223_{-2}^{+2}$ & $163.3_{-0.8}^{+0.8}$ \\
2007994281452374784 & 732.0 & 1.351 & $3552_{-4}^{+5}$ & $104.1_{-0.9}^{+0.8}$ \\
4848874011097672576 & 744.4 & 0.056 & $3727_{-22}^{+20}$ & $97.2_{-2.5}^{+2.9}$ \\
2526490979747026944 & 746.3 & 0.138 & $3318_{-3}^{+3}$ & $128.5_{-0.6}^{+0.6}$ \\
1664981631784328960 & 747.2 & 0.052 & $3648_{-9}^{+8}$ & $91.2_{-1.1}^{+1.1}$ \\
\enddata
\end{deluxetable*}